\documentclass[aps, prd, amsmath, amssymb, amsfonts, floats, %
floatfix, superscriptaddress, nofootinbib, twocolumn, showpacs]%
{revtex4}

\allowdisplaybreaks[2]
  
\usepackage{graphicx}
\usepackage[usenames, dvipsnames]{color}
\usepackage[colorlinks, pdfborder={0 0 0}, plainpages=false]{hyperref}
\usepackage{breakurl}
\usepackage{amsmath, amssymb, amsfonts}
\usepackage{xspace} 
\usepackage{dcolumn}
\usepackage{bm}
\usepackage{multirow} 

\definecolor{CiteColor}{rgb}{0, 0.5, 0}
\hypersetup{citecolor=CiteColor} %
\definecolor{RefColor}{rgb}{0.55, 0, 0}
\hypersetup{linkcolor=RefColor} %

\usepackage{ulem}
\usepackage{amsmath}
\definecolor {darkgreen}{rgb}{0.2, 0.7, 0.2}

\newcommand{\Maryland}{\affiliation{Maryland Center for Fundamental
    Physics \& Joint Space-Science Institute,\\ Department of Physics,
    University of Maryland, College Park, MD 20742, USA}}
\newcommand{\Caltech}{\affiliation{Theoretical Astrophysics 350-17,
    California Institute of Technology, Pasadena, CA 91125, USA}}
\newcommand{\Cornell}{\affiliation{Center for Radiophysics and Space
    Research, Cornell University, Ithaca, New York 14853, USA}}
\newcommand{\CITA}{\affiliation{Canadian Institute for Theoretical
    Astrophysics, 60 St.~George Street, University of Toronto,
    Toronto, ON M5S 3H8, Canada}} %
\newcommand{\CIFAR}{\affiliation{Canadian Institute for Advanced Research, 180 Dundas St.~West, Toronto, ON M5G 1Z8, Canada}} %

\newcommand{\vL}{\mbox{\boldmath${L}$}}
\newcommand{\vS}{\mbox{\boldmath${S}$}}
\newcommand{\vJ}{\mbox{\boldmath${J}$}}

\newcommand{\vN}{\mbox{\boldmath${N}$}}

\newcommand{\vvr}{\mbox{\boldmath${r}$}}
\newcommand{\vp}{\mbox{\boldmath${p}$}}
\newcommand{\ve}{\mbox{\boldmath${e}$}}
\newcommand{\vchi}{\mbox{\boldmath${\chi}$}}
\newcommand{\vlambda}{\mbox{\boldmath${\lambda}$}}

\newcommand{\vvOmega}{\mbox{\boldmath${\Omega}$}}

\begin{document}

\title{Inspiral-merger-ringdown waveforms of spinning, precessing black-hole binaries in the effective-one-body formalism}

\author{Yi Pan} \Maryland %
\author{Alessandra Buonanno} \Maryland %
\author{Andrea Taracchini} \Maryland %
\author{Lawrence E. Kidder} \Cornell 
\author{Abdul H.~Mrou\'{e}} \CITA 
\author{Harald P.~Pfeiffer} \CITA \CIFAR 
\author{Mark A.~Scheel} \Caltech 
\author{B\'{e}la Szil\'{a}gyi} \Caltech 

\begin{abstract}
We describe a general procedure to generate spinning, precessing waveforms that include inspiral, merger and 
ringdown stages in the effective-one-body (EOB) approach. The procedure uses a precessing frame in 
which precession-induced amplitude and phase modulations are minimized, and an inertial frame, 
aligned with the spin of the final black hole, in which we carry out the matching of the inspiral-plunge 
to merger-ringdown waveforms. As a first application, we build spinning, precessing EOB waveforms for the gravitational modes $\ell =2$ such that in the nonprecessing limit those 
waveforms agree with the EOB waveforms recently calibrated to numerical-relativity waveforms. 
Without recalibrating the EOB model, we then compare EOB and post-Newtonian precessing waveforms 
to two numerical-relativity waveforms produced by the Caltech-Cornell-CITA collaboration. The numerical waveforms are strongly 
precessing and have 35 and 65 gravitational-wave cycles. We find a remarkable agreement 
between EOB and numerical-relativity precessing waveforms and spins' evolutions. 
The phase difference is $\sim 0.2$ rad at merger, while the mismatches, computed 
using the advanced-LIGO noise spectral density, are below $2\%$ when maximizing only on the time and phase at coalescence and on the polarization angle.  
\end{abstract}

\date{\today}


\maketitle

\section{Introduction}
\label{sec:introduction}

An international network of gravitational-wave (GW) detectors
operating in the frequency band $10\mbox{--}10^3$ Hz exists today. It
is composed of the LIGO detectors in Hanford, WA
  and Livingston, LA, in the United States, the French-Italian Virgo
detector~\cite{Acernese:2008zz}, and the British-German GEO600
detector~\cite{Grote:2008zz}. Those detectors have collected and
analysed data for several years. Since 2010 they have been shut down
to be upgraded to the advanced LIGO and Virgo
configurations~\cite{Shoemaker2009}.  The design sensitivity for
advanced detectors, which is planned to be achieved by
2019~\cite{Aasi:2013wya}, will be a factor of ten better than the one
of the initial detectors. This improvement implies an increase in the
event rates of coalescing binary systems of (roughly) one thousand,
thus making very likely the first detection of gravitational
waves~\cite{Aasi:2013wya} with the advanced
  detector network. Furthermore, efforts to build a
gravitational-wave detector in space are
underway~\cite{ESALISAwebsite}.

Binary systems composed of compact objects, such as black holes and neutron stars (compact 
binaries for short) are the most promising sources for groundbased GW detectors. The signal detection 
and interpretation is based on a matched-filtering technique, where the
noisy detector output is cross-correlated with a bank of theoretical templates. 

Fueled by numerical relativity (NR) simulations, there has been substantial progress in building 
and validating accurate templates for the inspiral, merger and 
ringdown stages of nonprecessing~\footnote{Here, nonprecessing means that the BH spins are 
either zero or aligned/antialigned with the orbital angular momentum.} 
black-hole (BH) binaries~\cite{Buonanno2007, DN2007b, DN2008,Buonanno:2009qa,
Damour2009a, Pan:2009wj, PanEtAl:2011, Taracchini2012,Damour:2012ky, Ajith-Babak-Chen-etal:2007, Ajith:2008,
Ajith:2009bn,Santamaria:2010yb} 
(see also Ref.~\cite{Hinder:2013oqa} where several analytical templates 
have been compared to simulations produced by the NRAR collaboration). Despite this progress, 
template modeling for generic, spinning BH binaries is far from being complete.
 In this paper we focus on BH binary systems 
moving on quasi-spherical orbits, i.e., generic precessing orbits that are circularized and shrunk 
by radiation reaction. 

During the last several years, the post-Newtonian (PN) formalism,
which expands the Einstein equations in powers of $v/c$ ($v$ being the
characteristic velocity of the binary and $c$ the speed of light), has 
extended the knowledge of the dynamics and gravitational waveform for 
quasi-spherical orbits through 3.5PN~\cite{Blanchet:2011zv,Bohe:2013cla} and
2PN~\cite{Buonanno:2012rv} order, respectively. Precession-induced
modulations in the phase and amplitude of gravitational waveforms
become stronger as the opening angle between the orbital angular
momentum and the total angular momentum of the binary
increases. Compact binaries with large mass ratios can have 
larger opening angles than those with comparable masses. 

Pioneering studies aimed at understanding and
modeling precession effects in inspiraling compact binaries were
carried out in the mid 90s~\cite{Apostolatos1994,Apostolatos:1995pj}.
As GW detectors came online in early 2000, it became
more urgent to develop template families for spinning, precessing
binaries in which precession-induced modulations were incorporated in
an efficient way, reducing also the dimensionality of the parameter
space.  In 2003, Buonanno, Chen and Vallisneri~\cite{Buonanno:2002fy}
introduced the {\it precessing convention} and proposed a template
family for precessing binaries in which precessional effects are 
neatly disentangled from nonprecessing effects in both the amplitude
and phase evolutions. The precessing convention was initially
introduced to model phenomenological or detection template
families~\cite{Buonanno:2002fy}, and then it was extended to physical
templates of single-spin binary systems in
Refs.~\cite{Pan:2003qt,Buonanno:2005pt}.  In the past few years,
geometric methods have been developed to construct preferred precessing
reference frames~\cite{Schmidt2010,OShaughnessy2011,Boyle2011,Boyle:2013nka} 
for numerical or analytical waveforms, achieving a similar disentanglement 
of precessional effects. Waveforms decomposed in such frames exhibit 
relatively smooth amplitude and phase evolutions and are well 
approximated by nonprecessing waveforms~\cite{Schmidt:2012rh,Pekowsky:2013ska}.

Here, we use the effective-one-body (EOB) formalism~\cite{Buonanno99,Buonanno00,2000PhRvD..62h4011D,Damour01c} 
to model precessing inspiral, merger and ringdown waveforms. The basic idea of the EOB approach is to map by a canonical transformation 
the conservative dynamics of two compact objects of masses $m_1$ and $m_2$ and spins 
$\vS_1$ and $\vS_2$ into the dynamics of an effective particle of mass $\mu =
m_1\,m_2/(m_1+m_2)$ and spin $\vS_*$ moving in a deformed Kerr metric
with mass $M =m_1+m_2$ and spin $\vS_\text{Kerr}$, the deformation parameter being 
the symmetric mass ratio $\nu = \mu/M$. In the mid 2000s, Buonanno, Chen and Damour~\cite{Buonanno:2005xu} 
modeled EOB inspiral, merger and ringdown waveforms including for the first time 
spinning, precessing effects. 

In this paper we build on Refs.~\cite{Buonanno:2002fy,Buonanno:2005xu}, 
and also on the most recent analytical work~\cite{Barausse:2009aa,Barausse:2009xi,Barausse:2011ys} and the work 
at the interface between numerical and analytical relativity~\cite{Pan:2009wj,Taracchini2012}, and develop a 
general procedure to generate spinning, precessing waveforms in the EOB approach. The procedure employs 
the precessing convention of Ref.~\cite{Buonanno:2002fy} and an inertial frame aligned with the spin of 
the final BH. As a first application, we construct spinning, precessing waveforms that contain only the $\ell =2$ gravitational 
mode and reduce to the nonprecessing waveforms calibrated to numerical-relativity (NR) waveforms~\cite{Scheel2009,Boyle2007,Chu2009,Buchman:2012} in Ref.~\cite{Taracchini2012}. We compare these EOB precessing waveforms to Taylor-expanded PN waveforms and to two NR waveforms recently produced by the Caltech-Cornell-CITA collaboration~\cite{Mroue:2013xna}.  

The paper is organized as follows. In Sec.~\ref{sec:model-precession} we discuss the main coordinate 
frames that are used to describe precessing waveforms and their physical characteristics. 
We also review different proposals that have been suggested in the analytical and 
numerical-relativity communities for the precessing source frame, in which precession-induced 
modulations in the waveform's phase and amplitude are minimized. 
We also study how spin components parallel to the orbital plane affect the energy flux and multipolar waveforms. 
In Sec.~\ref{sec:EOB} we build EOB precessing waveforms using a precessing source frame aligned with the 
Newtonian orbital angular momentum and an inertial frame aligned with the 
direction of the final BH spin. In Sec.~\ref{sec:comparison} we compare EOB precessing waveforms computed 
in different precessing source frames and carry out comparisons between EOB, Taylor-expanded PN and NR 
precessing waveforms. Section~\ref{sec:conclusions} summarizes our main conclusions and future work.

\section{Modeling precessing waveforms}
\label{sec:model-precession}

\subsection{Conventions and inertial frames}
\label{sec:frames}

Throughout the paper, we adopt geometric units $G=c=1$ and the
Einstein summation convention, unless otherwise specified. The masses
of the BHs are $m_1$ and $m_2$ and we choose $m_1\ge
m_2$. The total mass, mass ratio and symmetric mass ratio are
$M=m_1+m_2$, $q= m_1/m_2$ and $\nu=m_1m_2/M^2$, respectively. The
position, linear momentum and spin vectors of the BHs are $\vvr_i(t)$,
$\vp_i(t)$ and $\vS_i(t)=\chi_i m_i^2\hat{\vS}_i$, where $i=1$, $2$
and $\chi_i$ is the dimensionless spin magnitude. In the EOB
framework, we solve the time evolution of the relative (rescaled) position vector 
$\vvr(t)\equiv\left(\vvr_1(t)-\vvr_2(t)\right)/M$, the center-of-mass--frame (rescaled) momentum vector 
$\vp(t)\equiv\vp_1(t)/\mu=-\vp_2(t)/\mu$, and the spins variables $\vS_1(t)$ and $\vS_2(t)$.

We start with an arbitrary orthonormal basis $\{\ve_x,\ve_y,\ve_z\}$. Without loss of generality,
we align the initial relative position vector $\vvr_0$ with $\ve_x$
and the initial orbital orientation
$[\hat{\vL}_N]_0\equiv\hat{\vL}_N(0)\equiv
\vvr_0\times\dot{\vvr}_0/|\vvr_0\times\dot{\vvr}_0|$ with $\ve_z$,
where we indicate with an over-dot the time derivative and
$\dot{\vvr}_0=\dot{\vvr}(0)$ is the initial relative velocity. The
initial spin directions are specified by the spherical coordinates
$\theta^S_1$, $\phi^S_1$, $\theta^S_{2}$ and $\phi^S_2$ associated
with this basis.

In the nonprecessing case, $\hat{\vL}_N$ is a constant and it is natural to choose a {\it source frame} described by the (unit) basis vectors $\{\ve_1^S,\ve_2^S,\ve_3^S\}$ whose $\ve_3^S$ is aligned with $\hat{\vL}_N$. In the precessing case, we choose to align the source frame $\{\ve_1^S,\ve_2^S,\ve_3^S\}$ with $\{\ve_x,\ve_y,\ve_z\}$. The basis vector $\ve_3^S$ is aligned with the initial orbital orientation $\hat{\vL}_N(0)$ but not with $\hat{\vL}_N(t)$ at later times.

\begin{figure}
  \begin{center}
    \includegraphics*[width=0.45\textwidth]{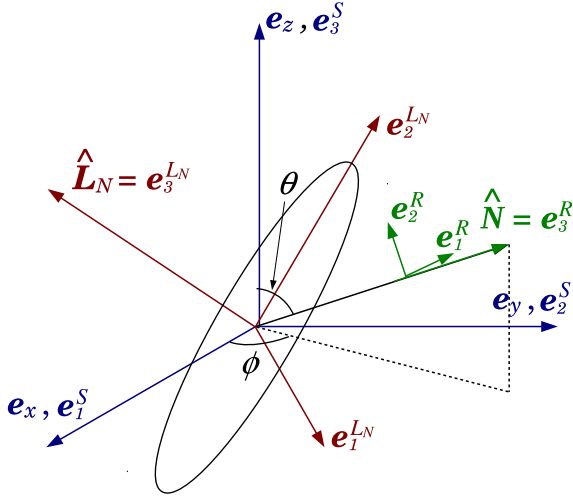}
    \caption{\label{fig:frames} We show the {\it radiation frame}
      $\{\ve_1^R,\ve_2^R,\ve_3^R\}$, the inertial {\it source frame}
      $\{\ve_1^S,\ve_2^S,\ve_3^S\}$ and the {\it precessing source frame}
      $\{\ve_1^{L_N},\ve_2^{L_N},\ve_3^{L_N}\}$ employed to describe a
      precessing BH binary and its GW radiation.} 
  \end{center}
\end{figure}

The GW polarizations $h_+$ and $h_\times$ can be obtained by 
projecting the strain tensor $h_{ij}$ onto the {\it radiation frame} described 
by the (unit) basis vectors $\{\ve_1^R,\ve_2^R,\ve_3^R\equiv\hat{\vN}\}$, the basis vector $\hat{\vN}$ 
being along the direction of propagation of the wave (see Fig.~\ref{fig:frames}). 
That is   
\begin{subequations}
\begin{eqnarray}
h_+      &=& \frac{1}{2}\left[\ve_1^R\otimes\ve_1^R-\ve_2^R\otimes\ve_2^R\right]^{ij}h_{ij} \,,\\
h_\times &=& \frac{1}{2}\left[\ve_1^R\otimes\ve_2^R+\ve_2^R\otimes\ve_1^R\right]^{ij}h_{ij} \,,
\end{eqnarray}
\end{subequations}
where the basis vectors $\ve_1^R$ and $\ve_2^R$ are defined by (see Fig.~\ref{fig:frames})
\begin{subequations}
\begin{eqnarray}
\ve_1^R &\equiv& \frac{\ve_z\times\hat{\vN}}{|\ve_z\times\hat{\vN}|} \,,\\
\ve_2^R &\equiv& \hat{\vN}\times\ve_1^R \,.
\end{eqnarray}
\end{subequations}
In the source frame $\{\ve_1^S,\ve_2^S,\ve_3^S\}$, we can decompose the polarizations $h_+$ and $h_\times$ in $-2$ spin-weighted spherical harmonics 
${}_{-2}Y_{\ell m}(\theta,\phi)$ as
\begin{equation}\label{hphc}
h_+(\theta,\phi) - ih_\times(\theta,\phi) = \sum_{\ell=2}^\infty\sum_{m=-\ell}^\ell {}_{-2}Y_{\ell m}(\theta,\phi)h_{\ell m}\,.
\end{equation}
The modes $h_{\ell m}$ can be calculated by applying the orthogonality condition valid for the ${}_{-2}Y_{\ell m}(\theta,\phi)$'s. 
Thus
\begin{equation}\label{hlm}
h_{\ell m}=\int\left[h_+(\theta,\phi) - ih_\times(\theta,\phi)\right]{}_{-2}Y_{\ell m}^*(\theta,\phi)\,d\Omega\,,
\end{equation}
where $\theta$ and $\phi$ are the inclination and azimuthal angles of the unit vector $\hat{\vN}$ as measured in the source frame.
Note that in the above expressions, we omit the dependence of the GW polarizations on time 
and binary parameters.

\subsection{Precessing source frames in analytical-relativity waveforms}
\label{sec:precessing-frame-AR}

In the nonprecessing case, the orbital orientation is constant
  and coincides with the directions of the orbital angular
  momentum $\vL \equiv \mu M \vvr \times \vp$, the Newtonian angular momentum
  $\vL_N \equiv \mu M \vvr \times \dot{\vvr}$, and the total angular
  momentum $\vJ \equiv \vL + \vS_{1}+\vS_{2}$. If the source frame is
aligned with the orbital orientation, the gravitational polarizations
are quite simple and are described mainly by the $(2,2)$ mode and a
few subdominant modes~\cite{PanEtAl:2011,Taracchini2012}. In this case, the wave's amplitude and
frequency increase monotonically during the inspiral and plunge
stages, and the amplitudes of the GW modes display a clean hierarchy. By
contrast, precessing waveforms decomposed in an inertial source frame
show strong amplitude and phase modulations. In this case the amplitudes of the 
GW modes do not necessarily follow a clean hierarchy~\cite{Arun:2008kb}.
Ideally we would like to conduct comparisons and calibrations between
analytical and numerical waveforms in a time dependent source frame that
minimizes precession-induced modulations. Fortunately, this is
possible if we choose a source frame that precesses with the binary
orbital plane~\cite{Buonanno:2002fy,Buonanno:2005pt,Schmidt2010,OShaughnessy2011,Boyle2011}.

Buonanno, Chen and Vallisneri~\cite{Buonanno:2002fy} proposed 
the {\it precessing convention} that neatly disentangles precessional effects 
from both amplitude and phase evolutions in restricted (i.e., leading-order) PN
waveforms. In the precessing convention, the precessing waveform is written as the product of a nonprecessing carrier waveform and a modulation term 
that collects all precessional effects.
In Refs.~\cite{Buonanno:2002fy,Buonanno:2005pt} the 
authors chose the precessing source frame aligned
with the Newtonian orbital angular momentum $\vL_N$. 
In this case, the basis vectors of the precessing source frame, $\{\ve_1^{L_N}(t),\ve_2^{L_N}(t),\ve_3^{L_N}(t)\}$ in Fig.~\ref{fig:frames}, read~\cite{Buonanno:2002fy} 
\begin{subequations}
\begin{eqnarray}
\ve_3^{L_N}(t)&=&\hat{\vL}_N(t) \,,\label{e3}\\
\dot{\ve}_1^{L_N}(t)&=&\vvOmega_e(t)\times\ve_1^{L_N}(t) \,,\label{e1}\\
\dot{\ve}_2^{L_N}(t)&=&\vvOmega_e(t)\times\ve_2^{L_N}(t) \,,\label{e2}
\end{eqnarray}
\end{subequations}
where
\begin{equation}\label{Omegae}
\vvOmega_e(t)\equiv\vvOmega_L(t)-\left[\vvOmega_L(t)\cdot\hat{\vL}_N(t)\right]\hat{\vL}_N(t)=\hat{\vL}_N(t)\times\dot{\hat{\vL}}_N(t) \,,
\end{equation}
and $\vvOmega_L(t)$ is the angular velocity of the precession of $\hat{\vL}_N(t)$ and 
satisfies $\dot{\hat{\vL}}_N(t)=\vvOmega_L(t)\times\hat{\vL}_N(t)$. Aligning the
precessing source frame with $\hat{\vL}_N(t)$ in Eq.~\eqref{e3}
removes the precession-induced amplitude modulations. Intuitively,
Eqs.~\eqref{e1}--\eqref{Omegae} impose that $\ve_1^{L_N}(t)$ and
$\ve_2^{L_N}(t)$ follow the precession of
$\ve_3^{L_N}(t)=\hat{\vL}_N(t)$, but do not precess around it. The key
point of the precessing convention is the removal of all
precession-induced modulations from the orbital phase $\Phi(t)$, so
that $\Phi(t)$ is simply given by the integral of the (monotonic)
orbital frequency $\Omega$, i.e. $\Phi(t)=\int\Omega(t') dt'$ (see for
details Sec.~IVA in Ref.~\cite{Buonanno:2002fy}). The freedom of
choosing the constant of integration, or the initial phase, is
degenerate with the only degree of freedom left in defining the
precessing source frame, namely a constant rotation of $\ve_1^{L_N}$
and $\ve_2^{L_N}$ around $\ve_3^{L_N}$.

We want to test now the precessing convention on inspiraling PN waveforms 
computed beyond the restricted approximation, i.e., beyond leading order. 
We employ the waveforms of Ref.~\cite{Arun:2008kb} that have spin-amplitude corrections 
through 1.5PN order. We decompose the 
$h_{\ell m}$'s in two source frames: (i) the inertial source frame aligned with the initial
total angular momentum $\vJ_0$~\cite{Arun:2008kb} and (ii) the precessing
source frame $\{\ve_1^{L_N}(t),\ve_2^{L_N}(t),\ve_3^{L_N}(t)\}$ defined
by Eqs.~\eqref{e3}--\eqref{Omegae}. The waveforms decomposed in the $\vJ_0$-frame 
are given explicitly in Appendix B of Ref.~\cite{Arun:2008kb}. 
We calculate the waveforms decomposed in the precessing $\vL_N(t)$-frame
from the waveforms decomposed in the $\vJ_0$-frame by 
properly rotating the $h_{\ell m}$ modes. 

In general, given a set of spin-weighted spherical harmonics $h_{\ell m}^{(o)}$ 
decomposed in an original 
frame and the Euler angles $(\alpha, \beta, \gamma)$ that define the rotation from the
original frame to a new frame, the modes $h_{\ell m}^{(n)}$ decomposed in the
new frame are given by~\cite{Arun:2008kb,Goldberg1967}
\begin{equation}\label{rotation}
h_{\ell m}^{(n)} = \sum_{m'=-\ell}^\ell {D^{\ell\;*}_{m'm}}(\alpha,\beta,\gamma) h_{\ell m'}^{(o)} \,,
\end{equation}
where ${D^{\ell\;*}_{m'm}}(\alpha,\beta,\gamma)$ is the complex conjugate of the Wigner $D$-matrix 
\begin{equation}\label{WignerD}
D^\ell_{m'm}(\alpha,\beta,\gamma)=(-1)^{m'}\sqrt{\frac{4\pi}{2\ell+1}}\,_{-m'}Y_{\ell m}(\beta,\alpha)e^{im'\gamma} \,,
\end{equation}
where $_{-m'}Y_{\ell m}$ is the spherical harmonic of spin-weight $-m'$. 
The transformation is closed among modes with the same index $\ell$. 
In this paper, we focus on the $\ell=2$ modes both for simplicity and because 
even when precession is present the $\ell=2$ modes still dominate. 
Nevertheless, the $\ell>2$ modes are not negligible and we plan to extend the 
precessing EOB model to those modes in the future, following the same approach 
we propose and demonstrate here with the $\ell=2$ modes. 

\begin{figure}
  \includegraphics[width=0.45\textwidth]{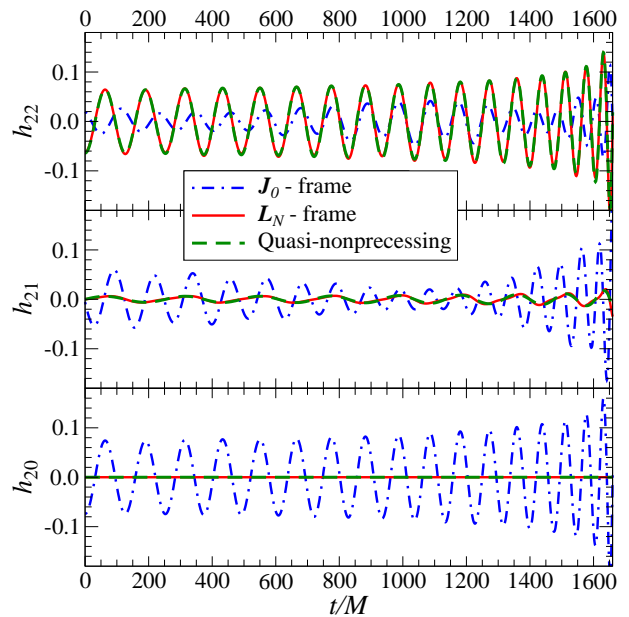}
  \caption{ \label{fig:PNwaves} We show inspiraling, precessing PN waveforms 
   decomposed in the inertial source frame aligned with the
    initial total angular momentum $\vJ_0$ and in the precessing source
    frame aligned with the Newtonian orbital angular momentum
    $\vL_N(t)$. For comparison, we show also the quasi-nonprecessing 
    PN waveforms defined in Sec.~\ref{sec:precessing-frame-AR}. The
    three panels use the same scale on the $y$-axis so that the amplitudes of the modes $h_{22}$,
    $h_{21}$ and $h_{20}$ can be easily compared.}
\end{figure}

In Fig.~\ref{fig:PNwaves}, we compare the $\vJ_0$-frame and
$\vL_N(t)$-frame $h_{2m}$ inspiraling waveforms emitted by a BH 
binary with mass ratio $q=6$ and spin magnitudes $\chi_1=\chi_2=0.8$. 
We choose spin orientations that give strong precession-induced modulations.  
As we can see, there is a clear hierarchy 
among the $h_{2m}$'s amplitudes  when decomposed in the $\vL_N(t)$-frame,  
but not when the decomposition is done in the $\vJ_0$-frame. In fact, 
in the $\vJ_0$-frame, the $(2,1)$ and $(2,0)$ modes have even larger amplitudes 
than the $(2,2)$ mode. We notice that the strong amplitude and phase modulations
of the modes in the $\vJ_0$-frame almost completely disappear when 
the $\vL_N(t)$-frame is used. 

Furthermore, we find it interesting to compare the modes of the
precessing waveforms to the ones of a ``nonprecessing'' binary system
having the same mass ratio and $\chi_i(t)\equiv\chi_i\hat{\vS}_i(t)
\cdot\hat{\vL}_N(t)\,,\;(i=1,2)$.  That is to say, we keep only the
components of the spin vectors along the direction of the Newtonian
angular momentum $\vL_N(t)$, and set all the other components to
zero. For convenience, we refer to such ``nonprecessing'' waveforms as the {\it
  quasi-nonprecessing} waveforms. We use the
adjective ``quasi'' because, differently from the nonprecessing waveforms, 
where the spins are aligned or antialigned with the orbital angular
momentum and remain constant throughout the evolution, in the
quasi-nonprecessing waveforms the spins evolve (according to
Eqs.~(\ref{EOM3})--(\ref{EOM4}) given below) and over time change their
projections onto $\vL_N$.  As can be seen in
Fig.~\ref{fig:PNwaves}, the near-perfect agreement between
$\vL_N(t)$-frame and quasi-nonprecessing waveforms indicates that the
spin components along $\vL_N(t)$ dominate the spin effects.  This
conjecture is reinforced by the observation that, because of parity
symmetry, the spin-orbit couplings contribute to the GW energy flux to infinity (known through
3.5PN order) only through terms of the form
$\vS_i(t)\cdot\hat{\vL}_N(t)\,,\;(i=1,2)$~
\cite{Kidder:1995zr,Blanchet-Buonanno-Faye:2006,Blanchet:2011zv,Bohe:2013cla}. The energy flux is a frame independent quantity. It is
given, in the adiabatic assumption, by~\footnote{Following the nonprecessing EOB model of
  Ref.~\cite{Taracchini2012}, we include in $dE/dt$ the spin-orbit terms 
  through 1.5PN order~\cite{Pan2010hz}, even if some of us have recently 
  computed the spin-orbit and spin-spin terms in the factorized flux 
  through 2PN order using results in Ref.~\cite{Buonanno:2012rv}.}
\begin{equation}\label{Edot}
  \frac{dE}{dt}=\frac{(M\Omega)^2}{8\pi}\sum_{\ell=2}^\infty\sum_{m=-\ell}^{\ell}m^2\left|\frac{\mathcal{R}}{M}h_{\ell m}\right|^2\,,
\end{equation}
where $\mathcal{R}$ is the distance to the source (and simply cancels 
the dependence on $\mathcal{R}$ hidden in the $h_{\ell m}$'s). The fact that the spin-orbit effects in
$dE/dt$ depend on spins only through $\vS_i(t)\cdot\hat{\vL}_N(t)\,,\;(i=1,2)$ suggests that the
dependence of the modes on spin's components parallel to the orbital plane 
disappears when all the modes are summed up to 
make the total energy flux. 

Therefore, beyond the leading-order results of Ref.~\cite{Buonanno:2002fy}, we find that 
PN precessing waveforms~\cite{Arun:2008kb} reduce to nearly nonprecessing waveforms when
decomposed to spin-weighted spherical harmonics in the source frame
$\{\ve_1^{L_N}(t),\ve_2^{L_N}(t),\ve_3^{L_N}(t)\}$ of the precessing convention~\cite{Buonanno:2002fy}. 
In addition, the PN quasi-nonprecessing waveforms are a good first
approximations to the $h_{\ell m}$'s modes in the precessing frame.

\subsection{Precessing source frames in numerical-relativity waveforms}
\label{sec:precessing-frame-NR}

The possibility of demodulating precessing waveforms using precessing source frames 
was recently verified with and generalized to NR 
waveforms in Refs.~\cite{Schmidt2010,Schmidt:2012rh,OShaughnessy2011}. In particular, 
Schmidt {\it et al.}~\cite{Schmidt2010,Schmidt:2012rh} and O'Shaughnessy {\it et al.}~\cite{OShaughnessy2011} 
identified the preferred {\it radiation axis} at infinity and showed that if a precessing 
frame aligned with the radiation axis is chosen, then the amplitude and phase modulations 
of numerical waveforms are removed and a clean hierarchy among the modes
is restored. 

In particular, Schmidt {\it et al.} proposed the so-called 
quadrupole-preferred frame in which the power of the $(\ell,\pm
m)=(2,2)$ mode is maximized. O'Shaughnessy {\it et al.} proposed a
more general and geometrical choice of the precessing frame in which the $z$
component of the radiated angular momentum is maximized. The latter proposal 
reduces to the choice of Schmidt {\it et al.} when the radiated angular momentum
is calculated using only the $(\ell,\pm m)=(2,2)$ modes. Boyle {\it et
  al.}~\cite{Boyle2011} then proposed the {\it minimal rotation}
condition to remove the remaining arbitrariness in the azimuthal
rotation of the precessing frame and in the phase modulations of 
the waveform. Given an inertial frame $\{\ve_x,\ve_y,\ve_z\}$ and
the first two Euler angles $\alpha(t)$ and $\beta(t)$ that align
$\ve_z$ with the radiation axis, the minimal rotation condition on the
third Euler angle $\gamma(t)$ is given by
\begin{equation}\label{min-rot}
\dot{\gamma}(t)=-\dot{\alpha}(t)\cos\beta(t)\,.
\end{equation}
This condition is equivalent to Eqs.~\eqref{e1}--\eqref{Omegae} above 
on the evolution of $\ve_1^{L_N}(t)$ and $\ve_2^{L_N}(t)$. If $\alpha(t)$ and $\beta(t)$
are the first two Euler angles that align $\ve_z$ with $\hat{\vL}_N$, then $\gamma(t)-\gamma(t_0)$ 
is the angle by which $\ve_1^{L_N}(t)$ and $\ve_2^{L_N}(t)$ shall rotate in the instantaneous 
orbital plane, relative to their positions at a reference time $t_0$, to satisfy the precession 
convention. Recently, Boyle~\cite{Boyle:2013nka} proposed a geometric definition of 
the angular-velocity vector of a waveform,  
to determine a frame in which the modes' amplitudes become very simple 
and the phases are nearly constant. 

Schmidt {\it et al.}~\cite{Schmidt:2012rh} showed that precessing PN inspiral 
  waveforms computed in the precessing source frame aligned with the preferred radiation axis
  are well approximated by nonprecessing PN waveforms. Furthermore, they
  proposed that precessing waveforms can be generated with
  good accuracy by transforming nonprecessing waveforms from
  precessing source frames to inertial source frames.  In a recent study,
  Pekowsky {\it et al.}~\cite{Pekowsky:2013ska} studied the mapping of precessing waveforms to
  nonprecessing ones using a large number of (short) numerical
  simulations and the analytical \verb+IMRPhenomB+~\cite{Ajith:2009bn} waveforms.
They found that precessional degrees of freedom that cannot be reproduced
by nonprecessing models (such as spin's components perpendicular 
to $\hat{\vL}_N$) give rise to corrections to the
nonprecessing waveforms that are very small during inspiral, but 
they can become significant during merger and ringdown.

\subsection{Strategy to build precessing effective-one-body waveforms}
\label{sec:strategy}

Motivated by the results discussed in Secs.~\ref{sec:precessing-frame-AR} and 
\ref{sec:precessing-frame-NR} of a nearly complete 
separation of precession-induced modulations in precessing waveforms when
using appropriate precessing source frames, we propose the following 
approach to generate generic EOB waveforms. 

First, we evolve the EOB dynamics and solve Eqs.~\eqref{e3}--\eqref{Omegae} 
for the precessing source frame
$\{\ve_1^{L_N}(t),\ve_2^{L_N}(t),\ve_3^{L_N}(t)=\hat{\vL}_N(t)\}$. Since
the difference between $\vL_N$ and $\vL$ starts at 1PN order, the
leading-order conclusions achieved by the precessing $\vL_N$-frame
hold if we replace $\vL_N$ with $\vL$ in Eqs.~\eqref{e3}--\eqref{Omegae}. We have verified that precessing
waveforms decomposed in the $\vL$-frame agree equally well with the
quasi-nonprecessing waveforms generated by keeping only spin's 
components along $\vL$. Furthermore, in
  Sec.~\ref{sec:comparison} we compare precessing EOB waveforms
  (generated either in the $\vL_N$-frame or in the $\vL$-frame) to NR
  waveforms, and find that their mutual difference is
  marginal. Without a more accurate calibration and comprehensive
  comparisons with NR waveforms, we do not know {\it a priori} whether
  the $\vL_N$-frame or the $\vL$-frame is more preferable, nor can we
  say which of them captures the precession effects more faithfully. 
Thus, at the current stage, we simply adopt the $\vL_N$-frame as the default
precessing source frame in the EOB model. 

Second, because of the simple features of the inspiral-plunge modes in
the precessing source frame --- little modulation and clean hierarchy
--- we choose to model the precessing inspiral-plunge EOB modes in
this frame, and generate modes in any arbitrary source frame through
Eq.~\eqref{rotation}. Since factorized EOB modes for precessing spins 
are not available yet and EOB modes have been calibrated only to 
nonspinning and spinning, nonprecessing NR modes~\cite{PanEtAl:2011,Taracchini2012}, 
we choose to work in the precessing source frame and use 
quasi-nonprecessing modes as good approximations to precessing modes 
(as discussed in Secs.~\ref{sec:precessing-frame-AR} and \ref{sec:precessing-frame-NR}). 
In particular, we employ the quasi-nonprecessing inspiral-plunge modes based on the latest
spinning, nonprecessing EOB model that was calibrated to NR modes in
Ref.~\cite{Taracchini2012}. Note that we are not obliged to use in the future 
quasi-nonprecessing waveforms in the precessing source frame. As soon
as factorized EOB modes for precessing spins become
available, we shall relax the assumption of using quasi-nonprecessing
inspiral-plunge modes~\footnote{It remains to be investigated, though, 
whether it is necessary to include precessing effects in the EOB modes 
decomposed in the precessing source frame to meet more stringent 
accuracy requirements for advanced LIGO and Virgo searches.}. The strategy that we present
in this paper is generic and can easily be applied to future 
calibrations and analytical improvements of the EOB model. 

Third, we rotate the quasi-nonprecessing modes from the precessing
source frame to the inertial frame whose $z$-axis coincides with the
direction of the total angular momentum $\vJ$ at a time very close to
merger when the direction of $\vJ$ is a good approximation to the  
direction of the spin of the final BH.  In this inertial frame we
match the inspiral-plunge to merger-ringdown modes
following the usual prescription in the EOB approach~\cite{Taracchini2012}. After generating
inspiral-merger-ringdown modes in this frame, it is straightforward to
calculate EOB modes $h_{\ell m}$ in any source frame or EOB
polarizations $h_{+,\times}$ in any radiation frame.

\section{Precessing effective-one-body model}
\label{sec:EOB}

In this section we construct a generic, precessing EOB model following the 
general strategy outlined above --- 
it employs the precessing source frame introduced in Ref.~\cite{Buonanno:2002fy} 
and the quasi-nonprecessing waveforms based on the nonprecessing EOB model developed 
in Ref.~\cite{Taracchini2012}. 

\subsection{Effective-one-body dynamics}
\label{sec:EOB-dyn}

Since we employ exactly the same EOB dynamics calibrated against NR 
simulations in Ref.~\cite{Taracchini2012}, we review only the key ingredients of the 
dynamics and refer the readers to Ref.~\cite{Taracchini2012} for further details.

The EOB dynamics of spinning BH binary systems is obtained solving the following 
Hamilton equations 
\begin{subequations}
  \begin{eqnarray}\label{EOM1}
    \frac{d\vvr}{d\hat{t}}&=&\{\vvr,\hat{H}_{\text{real}}\}=\frac{\partial \hat{H}_{\text{real} }}{\partial \vp}\,,\\\label{EOM2}
    \frac{d\vp}{d\hat{t}}&=&\{\vp,\hat{H}_{\text{real}}\}+\hat{\bm{\mathcal{F}}}=-\frac{\partial \hat{H}_{\text{real}}}{\partial \vvr}
    +\hat{\bm{\mathcal{F}}}\,,\\\label{EOM3}
    \frac{d\vS_1}{dt} &=& \{\vS_1,\mu \hat{H}_{\text{real}} \} = \mu \frac{\partial \hat{H}_{\text{real}}}{\partial \vS_1} \times \vS_1 \,,\\\label{EOM4}
    \frac{d\vS_2}{dt} &=& \{\vS_2,\mu \hat{H}_{\text{real}} \} = \mu \frac{\partial \hat{H}_{\text{real}}}{\partial \vS_2} \times \vS_2 \,,
  \end{eqnarray}
\end{subequations}
where $\hat{t}\equiv t/M$ is the dimensionless time variable, 
$\hat{H}_{\text{real}}$ is the reduced EOB Hamiltonian derived
in Refs.~\cite{Barausse:2009aa,Barausse:2009xi,Barausse:2011ys} and
reviewed in Sec.~IIA of Ref.~\cite{Taracchini2012}, and
$\hat{\bm{\mathcal{F}}}$ is the reduced radiation reaction
force. Following Ref.~\cite{Buonanno:2005xu}, we use
\begin{equation}
  \hat{\bm{\mathcal{F}}}=\frac{1}{\nu \hat{\Omega} |\vvr\times
    \vp|}\frac{dE}{dt}\vp\,,
\end{equation}
where $\hat{\Omega}\equiv M |\vvr\times\dot{\vvr}|/r^2$ is the
dimensionless orbital frequency and $dE/dt$ is the energy flux for
quasi-spherical orbits. We use Eq.~\eqref{Edot} with $\ell \leq 8$, namely 
\begin{equation}\label{Edot8}
  \frac{dE}{dt}=\frac{\hat{\Omega}^2}{8\pi}\sum_{\ell=2}^8\sum_{m=-\ell}^{\ell}m^2\left|\frac{\mathcal{R}}{M}h_{\ell m}\right|^2\,.
\end{equation}
Because under a change of frame the $h_{\ell m}$ modes for a given $\ell$ 
transform into modes with the same $\ell$, Eq.~(\ref{Edot8}) is still frame-independent. 
We insert the quasi-nonprecessing modes $h_{\ell m}$, i.e., the modes decomposed in the precessing source 
frame aligned with $\vL_N(t)$, into Eq.~\eqref{Edot8}. 
The quasi-nonprecessing modes can be calculated directly in the inertial
frame $\{\ve_x,\ve_y,\ve_z\}$ in which we solve the EOB dynamics. The only 
difference from the procedure of Ref.~\cite{Taracchini2012} is to replace the constant $\chi_1$ and $\chi_2$ with
$\chi_1\hat{\vS}_1(t)\cdot\hat{\vL}_N(t)$ and $\chi_2\hat{\vS}_2(t)\cdot\hat{\vL}_N(t)$. 

\subsection{Initial conditions}
\label{sec:EOB-ICs}

For applications in data analysis and comparisons with numerical or
analytical waveforms, we need initial conditions that start
the orbital evolution with sufficiently small eccentricity at a given
orbital separation (or GW frequency) and spins orientation. The analytical
quasi-spherical initial conditions proposed in Ref.~\cite{Buonanno:2005xu}
is a good first approximation.

In the nonprecessing case \cite{Pan:2009wj, PanEtAl:2011,Taracchini2012}, 
the residual eccentricity can be further reduced by starting the evolution 
at a larger separation (smaller GW frequency) and waiting for the orbits to 
be better circularized by radiation reaction. In the precessing case, 
however, we can not easily reduce the eccentricity in this way because  
we need specific spin directions at the initial separation. In order 
to reduce eccentricity by starting the evolution at a larger separation, 
we need to figure out what are the spin directions at this larger separation 
to ensure the desired spin directions at a given (smaller) initial separation. 
To reach this goal and reduce the eccentricity for quasi-spherical initial conditions 
we employ the method developed in Ref.~\cite{Buonanno:2010yk}, which 
is based on~\cite{Pfeiffer-Brown-etal:2007,Boyle2007}
~\footnote{This
method has been employed to reduce eccentricity in NR simulations of BH 
binary systems~\cite{Chu2009,Lovelace:2010ne,Buchman:2012,Mroue:2012kv,Purrer:2012wy,Purrer:2013ojf}.}. 
Thus, we 
first evolve the binary for a few orbits and estimate the eccentricity through
oscillations in orbital frequency $\Omega$ and separation $r$. We then
apply corrections to the initial conditions following Eqs.~(74) and (75) in 
Ref.~\cite{Buonanno:2010yk}. We repeat these steps until the eccentricity is
sufficiently small.

\subsection{Nonprecessing effective-one-body waveforms}
\label{sec:EOB-np-wave}

The EOB nonprecessing inspiral-plunge modes $h_{\ell m}^{\text{NP, insp-plunge}}$ developed in Ref.~\cite{Taracchini2012} are given by
\begin{equation}
  h_{\ell m}^{\text{NP, insp-plunge}}=h^{\text{F}}_{\ell m}\,N_{\ell m}\,,
\end{equation}
where $h_{\ell m}^{\text{F}}$ are the factorized modes derived in
Refs.~\cite{Damour2007, DIN, Pan2010hz}, and $N_{\ell m}$ are 
nonquasicircular (NQC) corrections that model deviations from
the quasicircular motion, that is assumed when deriving $h_{\ell
  m}^{\text{F}}$. The factorized modes read
\begin{equation}
h^\mathrm{F}_{\ell m}=h_{\ell m}^{(N,\epsilon)}\,\hat{S}_\text{
    eff}^{(\epsilon)}\, T_{\ell m}\, e^{i\delta_{\ell
      m}}\left(\rho_{\ell m}\right)^\ell \,,
\end{equation}
where $\epsilon$ is the parity of the mode. All the factors
entering $h_{\ell m}^{\text{F}}$ can be explicitly found in the
Appendix of Ref.~\cite{Taracchini2012}. As discussed above, when using 
these expressions to generate quasi-nonprecessing modes, the 
only minor modification we have to take into account is to replace the
constant spin magnitudes $\chi_1$ and $\chi_2$ by their time dependent
counterparts. More specifically, in the nonprecessing case, the leading
order spin-orbit effects in $\rho_{\ell m}$ are parametrized by two
linear combinations of the constant dimensionless spin parameters
\begin{subequations}
\begin{eqnarray}
\chi_S\equiv\frac{\chi_1+\chi_2}{2} \,,\\
\chi_A\equiv\frac{\chi_1-\chi_2}{2} \,.
\end{eqnarray}
\end{subequations}
In the precessing case, both $\chi_S$ and $\chi_A$ become linear combinations of the time varying spin vectors projected along $\hat{\vL}_N(t)$,
\begin{subequations}
\begin{eqnarray}
\label{chiSt}
\chi_S(t)\equiv\frac{1}{2}\left(\frac{\vS_1(t)}{m_1^2}+\frac{\vS_2(t)}{m_2^2}\right)\cdot\hat{\vL}_N(t) \,,\\
\label{chiAt}
\chi_A(t)\equiv\frac{1}{2}\left(\frac{\vS_1(t)}{m_1^2}-\frac{\vS_2(t)}{m_2^2}\right)\cdot\hat{\vL}_N(t) \,.
\end{eqnarray}
\end{subequations}
In Ref.~\cite{Taracchini2012}, the inspiral-merger-ringdown 
mode $(2,2)$ was calibrated against NR simulations. 
Studies in the test-particle nonspinning~\cite{DIN} and 
spinning, nonprecessing~\cite{Pan2010hz} cases suggest 
that the factorized modes $h_{\ell m}^{\text{F}}$ 
are good approximations of the inspiral-plunge modes 
even without any NQC correction or calibration. 
Thus, we model the inspiral-plunge $(2,1)$ mode with 
$h_{21}^{\text{F}}$. The $(2,0)$ mode has been computed in 
PN theory at leading order and its amplitude is 
$5/14\sqrt{6}\simeq 0.15$ times the amplitude of the leading order 
$(2,2)$ mode~\cite{Kidder:2007rt}. However, this prediction does 
not agree with NR results. In fact, we find~\cite{mode20} that for the 
nonspinning NR simulations of mass ratios $q = 1,6$~\cite{PanEtAl:2011},
the amplitude of the $(2,0)$ mode during the inspiral is smaller than the 
one of the $(2,2)$ mode by a factor $\sim 10^{3}$. Since we do not yet understand 
the origin of this discrepancy we have decided that in this first investigation we 
neglect the nonprecessing EOB $(2,0)$ mode.

The NQC correction to the $(2,2)$ mode, $N_{22}$, is given by
\begin{equation}\label{NQC}
  \begin{split}
    N_{22} &= \Bigg[1 + \left( \frac{p_{r^*}}{r\,\hat{\Omega}}
    \right)^{\!2}\! \Bigg(a_1^{h_{22}}\! +\! \frac{a_2^{h_{22}}}{r}\! +\!
    \frac{a_3^{h_{22}}}{r^{3/2}} 
     \!+\!\frac{a_4^{h_{22}}}{r^{2}}\! +\! \frac{a_5^{h_{22}}}{r^{5/2}}
    \Bigg) \Bigg]\\
& \times \exp \Bigg[i \frac{p_{r^*}}{r\,\hat{\Omega}}
    \Bigg(b_1^{h_{22}} 
    + p_{r^*}^2 b_2^{h_{22}} \!+\! \frac{p_{r^*}^2}{r^{1/2}}
    b_3^{h_{22}}+ \frac{p_{r^*}^2}{r} b_4^{h_{22}} \Bigg) \Bigg],
  \end{split}
\end{equation}
where the amplitude coefficients $a_i^{h_{22}}$ (with $i=1...5$) and
the phase coefficients $b^{h_{22}}_i$ (with $i=1...4$) are obtained
through the iterative procedure described in Sec.~IIB of
Ref.~\cite{Taracchini2012}. Since only equal-mass, equal-spin,
nonprecessing NR simulations were used to calibrate the EOB model of
Ref.~\cite{Taracchini2012}, we have to map the $N_{22}$ from generic
spin configurations to equal-spin, nonprecessing
configurations. Without further
calibrations, we first adopt a mapping from
precessing to nonprecessing configurations that equates the
  $\chi_S(0)$ and $\chi_A(0)$ of a precessing configuration (defined
  in Eqs.~\eqref{chiSt} and \eqref{chiAt}) to the constant $\chi_S$
  and $\chi_A$ of a nonprecessing configuration. Then, we apply the mapping
from a generic nonprecessing configuration to an equal-spin,
nonprecessing configurations as defined in Sec.~IVA of
Ref.~\cite{Taracchini2012}.

\subsection{Precessing source frame}
\label{sec:p-frame}

Let $\ve_3^{L_N}(t)=\hat{\vL}_N(t)$ be the third (unit) basis vector of the
precessing source frame. We solve the other two (unit) basis vectors $\ve_1^{L_N}(t)$ and
$\ve_2^{L_N}(t)$ by applying the minimal rotation condition. We do it because 
the latter involves only one differential equation, namely Eq.~\eqref{min-rot},
instead of Eqs.~\eqref{e1} and \eqref{e2} for the precessing convention. Specifically, with the help
of the inertial source frame $\{\ve_x^S,\ve_y^S,\ve_z^S\}$, we define
\begin{subequations}
\begin{eqnarray}\label{alpha}
\alpha(t)&=&\arctan \left [\frac{\ve_3^{L_N}(t)\cdot\ve_y^S}{\ve_3^{L_N}(t)\cdot\ve_x^S} \right ]\,,\\
\beta(t)&=&\arccos \left [\ve_3^{L_N}(t)\cdot\ve_z^S \right ]\,,
\label{beta}
\end{eqnarray}
\end{subequations}
and solve~\footnote{Following Boyle {\it et al.}~\cite{Boyle2011}, we
  integrate $\gamma(t)$ by parts and implement
  $\gamma(t)=-\alpha(t)\cos\beta(t)-\int\alpha(t)\dot{\beta}(t)\sin\beta(t)\,dt$
  to avoid differentiating $\alpha(t)$, which can be noisy near the
  coordinate singularities at $\beta(t)=0$ and $\beta(t)=\pi$. We note that Boyle~\cite{Boyle:2013nka} recently proposed a much more accurate and robust method to integrate $\gamma(t)$ using quaternions.} 
\begin{equation}\label{min-rot-2}
\dot{\gamma}(t)=-\dot{\alpha}(t)\cos\beta(t) \,.
\end{equation}
Those Euler angles $\alpha(t)$, $\beta(t)$ and $\gamma(t)$
describe the time-dependent rotation from the inertial source frame
$\{\ve_x^S,\ve_y^S,\ve_z^S\}$ to the precessing source frame
$\{\ve_1^{L_N}(t),\ve_2^{L_N}(t),\ve_3^{L_N}(t)\}$ with the
latter satisfying the minimal rotation condition. The only freedom in
the definition of the precessing source frame is a constant shift in
$\gamma(t)$ that is degenerate with the initial orbital phase.

\subsection{Precessing effective-one-body waveforms}
\label{sec:EOB-p-wave}

We build the complete inspiral-plunge-merger-ringdown waveforms in an 
inertial frame following the usual procedure in the EOB approach~
\cite{Buonanno99, Buonanno00, Buonanno-Cook-Pretorius:2007,
  Buonanno2007, DN2007b, DN2008, Buonanno:2009qa, Damour2009a,
  Pan:2009wj, PanEtAl:2011, Taracchini2012, Damour:2012ky}. More 
specifically, we join the inspiral-plunge waveform $h_{\ell m}^{\text{insp-plunge}}$ and the
merger-ringdown waveform $h_{\ell m}^{\text{merger-RD}}$ at a matching
time $t^{\ell m}_{\rm match}$ as
\begin{equation}
  \begin{split}
    h^{\text{EOB}}_{\ell m}(t) &= h_{\ell m}^{\text{inspiral-plunge}}(t)\,
    \theta(t_{\text{match}}^{\ell m}-t) \\
    &\quad +h_{\ell m}^{\text{merger-RD}}(t)\,
    \theta(t-t_{\text{match}}^{\ell m}) \,.
  \end{split}
\end{equation}
Given the quasi-nonprecessing inspiral-plunge modes $h_{\ell
  m}^{\text{NP, insp-plunge}}$ decomposed in the precessing source
frame and the Euler angles (not necessarily those in
Eqs.~\eqref{alpha}--\eqref{min-rot-2}, which are specific to
$\{\ve_x^S,\ve_y^S,\ve_z^S\}$) defining the rotation from the
precessing source frame to any inertial frame, the inspiral-plunge
modes in the inertial frame are given by Eqs.~\eqref{rotation}. To
study the $h_{2m}$ modes, we need all $\ell=2$ modes in the precessing
source frame. As discussed in Sec.~\ref{sec:EOB-np-wave}, we employ
the calibrated $(2,2)$ mode of Ref.~\cite{Taracchini2012},
$h_{21}^{\text{F}}$ for the $(2,1)$ mode, and zero for the $(2,0)$
mode.  In the precessing source frame, since we use
quasi-nonprecessing inspiral-plunge modes to approximate precessing
modes, we further assume reflection symmetry, which, combined with
parity invariance, gives modes with $m<0$ through
$h_{2\,,-m}^{\text{NP, insp-plunge}}(t)=h_{2 m}^{\text{NP,
    insp-plunge}\,*}(t)$. Pekowsky {\it et
  al.}~\cite{Pekowsky:2013ska} discussed how this symmetry is broken
by precessional effects, giving rise to a contribution to the $(2,2)$
mode which is odd under reflection. In the only example investigated 
in Ref.~\cite{Pekowsky:2013ska}, the ratio between the component 
of the $(2,2)$ mode of the Weyl scalar 
$\Psi_{4,22}$ that is odd under reflection and the one 
that is even under reflection is $\sim 0.01$, while the ratio between 
the former and the $(2,1)$ mode of the Weyl scalar $\Psi_{4,21}$ is 
$\sim 1$. Since the Weyl scalar and the metric perturbation are 
related by $\Psi_{4,\ell m}=\ddot{h}_{\ell m}\simeq m^2\hat{\Omega}^2h_{\ell m}$ 
during the inspiral, the odd component of the $h_{22}$'s amplitude is about a fourth 
of the $h_{21}$'s. The odd component of the $h_{22}$'s amplitude becomes 
substantial only during the merger and ringdown. Thus, in this first 
study, we ignore the component of the $(2,2)$ mode that is odd 
under reflection when describing the inspiraling 
waveform in the precessing frame, but we include the odd component 
when building the merger-ringdown waveform.

It is convenient to choose an inertial frame in which the
merger-ringdown waveforms take simple forms. A natural choice is the
frame aligned with the spin of the final BH, in which the
merger-ringdown waveforms are expressed as linear combinations of the
quasinormal modes (QNMs)~\cite{Buonanno99, Buonanno00, Buonanno-Cook-Pretorius:2007,
  Buonanno2007, DN2007b, DN2008, Buonanno:2009qa, Damour2009a,
  Pan:2009wj, PanEtAl:2011, Taracchini2012, Damour:2012ky}. Barausse
{\it et al.}~\cite{Barausse2009} found strong evidence that the spin of
the final BH is aligned with the initial total angular momentum of the
binary. Using this assumption they derived accurate formulas for the 
final spin of a BH formed by merger. The success of their model verifies 
the PN-motivated assumption that the radiated angular momentum averaged 
over precessional cycles is almost aligned with the total angular momentum. 
Thus, the direction of the latter is preserved with high accuracy during 
the inspiral~\cite{Apostolatos1994}. Here we employ the formulas in Ref.~\cite{Barausse2009} 
to predict the magnitude of the spin of the final BH, and we 
align the final-spin direction with $\vJ(t^{\rm EOB}_{\rm \Omega peak})$, which 
is the total angular momentum at the time the EOB orbital frequency reaches 
its peak ($t^{\rm EOB}_{\rm \Omega peak}$). The time $t^{\rm
  EOB}_{\rm \Omega peak}$ has been adopted in most previous EOB models
as the reference time of merger \cite{Buonanno:2009qa, Damour2009a,
  Pan:2009wj, PanEtAl:2011, Taracchini2012, Damour:2012ky}. Without further 
information from NR simulations of precessing, spinning BHs, 
we consider ${\vJ}(t^{\rm EOB}_{\rm \Omega peak})$ our best prediction of 
the final-spin direction. We expect that not a lot of angular momentum is 
radiated during the swift transition from merger to ringdown~\cite{Buonanno-Cook-Pretorius:2007} 
and the small amount being radiated is likely to be nearly aligned with $\vJ(t^{\rm EOB}_{\rm \Omega peak})$. 

The inspiral-plunge waveform $h_{\ell m}^{\text{insp-plunge}}$ in the
inertial frame aligned with $\vJ(t^{\rm EOB}_{\rm \Omega peak})$
contains NQC corrections from the nonprecessing $(2,2)$ mode $h_{\ell
  m}^{\text{NP, insp-plunge}}$. Those corrections are derived based on
the assumption that the inspiral-plunge waveforms in the precessing
frame are the calibrated nonprecessing waveforms generated with the
specific mapping of spin parameters defined in
Sec.~\ref{sec:EOB-np-wave}. Although we expect that such assumption 
introduces systematic errors in $h_{\ell m}^{\text{insp-plunge}}$, we
are not able to quantify them before comparing $h_{\ell
  m}^{\text{insp-plunge}}$ with precessing NR waveforms. Therefore, we
do not apply any further correction to the inspiral-plunge waveform in
this model. This choice also guarantees that $h_{\ell
  m}^{\text{insp-plunge}}$ modes reduce to the calibrated modes of Ref.~\cite{Taracchini2012} 
in the nonprecessing limit.

The merger-ringdown waveform is built following almost exactly the approach
described in Ref.~\cite{Taracchini2012}. We first give a brief review of this approach and 
then describe the differences. The merger-ringdown waveform is modeled by a linear 
superposition of the QNMs of the final Kerr BH as
\begin{equation}\label{ringdown}
  h_{\ell m}^{\text{merger-RD}}(t)=\sum_{n=0}^{N-1}
  A_{\ell mn}\,e^{-i\sigma_{\ell mn}(t-t_{\text{match}}^{\ell m})}\,,
\end{equation}
where $N$ is the number of overtones, $A_{\ell mn}$ is the complex
amplitude of the $n$-th overtone of the $(\ell,m)$ mode, and
$\sigma_{\ell mn}$ is the complex frequency of the $n$-th
overtone. The complex frequencies are known function of the mass and
spin of the final BH~\cite{Berti2006a}. The mass of the final BH is given in Eq.~(8)
of Ref.~\cite{Tichy2008}. The spin magnitude of the final BH, as
discussed earlier, is given in Eqs.~(6), (8) and (10) of
Ref.~\cite{Barausse2009}. Following Ref.~\cite{Taracchini2012}, we
replace the highest physical overtone (the $7$-th) of the $(2,2)$ mode with
a pseudo QNM whose calibrated complex frequency is given in Eqs.~(35a)
and (35b) of Ref.~\cite{Taracchini2012}. Finally, we fix the complex
amplitudes $A_{\ell mn}$ though a matching procedure~\cite{PanEtAl:2011} 
that imposes a $C^1$-smooth connection over a 
time interval $\Delta t_{\rm match}^{\ell m}$ between the
merger-ringdown waveform and the inspiral-plunge waveform, in the
inertial frame aligned with $\vJ(t^{\rm EOB}_{\rm \Omega peak})$.

In Ref.~\cite{Taracchini2012}, the matching time
$t_{\text{match}}^{\ell m}$ and the time interval $\Delta t_{\rm
  match}^{\ell m}$ were calibrated only for the $(2,2)$ mode. 
Here we need to specify those quantities also for the remaining 
$\ell =2$ modes. We find that in order 
to keep the matching procedure stable when the binary is 
strongly precessing around merger, we have to introduce  
in  $t_{\text{match}}^{\ell m}$ and $\Delta t_{\rm match}^{\ell m}$ a
  dependence on how much the orbital and total angular momentum are
  misaligned at merger, i.e. on the quantity $\hat{\vL}(t^{\rm
    EOB}_{\rm \Omega peak})\cdot\hat{\vJ}(t^{\rm EOB}_{\rm \Omega
    peak})$. More specifically, in strongly precessing cases, the
  directions of $\vL(t)$ and $\vJ(t)$ can be very different close to
  merger. As a consequence, the inspiral-plunge modes in the inertial
  $\vJ(t^{\rm EOB}_{\rm \Omega peak})$-frame can present strong
  amplitude and frequency oscillations around merger. [Technically
  those strong oscillations are generated by drastic time-dependent
  rotations from well-behaved quasi-nonprecessing inspiral-plunge
  modes as the merger is approached.] Thus, to keep the matching
  procedure stable in strongly precessing situations we set the matching
  time $t_{\text{match}}^{\ell m}$ earlier and make the matching
  interval $\Delta t_{\rm match}^{\ell m}$ longer. We choose
\begin{eqnarray}
t_{\text{match}}^{\ell m} &=& t_{\text{match}}^{22,\rm Cal} - 10 M\left(1-|\kappa_{LJ}(t^{\rm EOB}_{\rm \Omega peak})|\right) \,, \\
\Delta t_{\rm match}^{\ell m} &=& \Delta t_{\rm match}^{22,\rm Cal}\left(10-9|\kappa_{LJ}(t^{\rm EOB}_{\rm \Omega peak})|\right) \,,
\end{eqnarray} 
where 
\begin{eqnarray}
t_{\text{match}}^{22,\rm Cal} &=& t^{\rm EOB}_{\rm \Omega peak}-
\begin{cases}
2.5M & \chi \le 0 \\
2.5M + 1.77M\left(\dfrac{\chi}{0.437}\right)^4 & \chi > 0
\end{cases} \,, \nonumber\\
\Delta t_{\rm match}^{22,\rm Cal} &=& 7.5M \nonumber\\
\end{eqnarray}
are the calibrated values of the $(2,2)$ mode in Ref.~\cite{Taracchini2012},
\begin{equation}
\chi=\chi_S+\chi_A\frac{\sqrt{1-4\nu}}{1-2\nu}
\end{equation}
is a linear combination of initial spin projections on $\vL_N$, and 
\begin{equation}
\kappa_{LJ}(t^{\rm EOB}_{\rm \Omega peak})=\hat{\vL}(t^{\rm EOB}_{\rm \Omega peak})\cdot\hat{\vJ}(t^{\rm EOB}_{\rm \Omega peak})
\end{equation}
is the cosine of the opening angle between the orbital and total
angular momenta at the reference time of merger $t^{\rm EOB}_{\rm
  \Omega peak}$.  When $\kappa_{LJ}(t^{\rm EOB}_{\rm \Omega
    peak})=0$, the matching time $t_{\text{match}}^{\ell m}$ is $10M$
  earlier than that of the aligned case, and the time interval $\Delta
  t_{\rm match}^{\ell m}$ is $10$ times that of the aligned case. The
  choice of $10M$ and the factor of $10$ made in this paper are rather
  arbitrary. They are based on the only requirement of producing qualitatively
  sound merger-ringdown waveforms.

\begin{table}
  \begin{ruledtabular}
    \begin{tabular}{c|cccccccc}
      Case & $q$ & $\chi_1$ & $\chi_2$ & $\theta_1^S$ & $\theta_2^S$ & $\phi_1^S$ & $\phi_2^S$ & $M\Omega_0$ \\
      \hline
      1 & $2$ & $0.6$ & $0.6$ & $\pi/3$ & $\pi/3$ & 0 & $\pi/2$ & 
$0.0112$ \\ 2 & $6$ & $0.8$~\footnote{The nonprecessing EOB model of
        Ref.~\cite{Taracchini2012} generates waveforms with any mass
        ratio and individual spin magnitudes $-1\le\chi_i\lesssim
        0.7$. Although we consider here $\chi_1=0.8$ because 
        $\chi_1(t)=\chi_1\hat{\vS}_1(t) \cdot \hat{\vL}_N(t) < 0.7$
        during the entire evolution, we do not find any problem in this case when generating
        the quasi-nonprecessing waveforms.}  & $0.6$ & $\pi/2$ & $2\pi/3$
        & $\pi/2$ & $\pi/2$ &
$0.0112$ \\
      3 & $3$ & $0.500$ & $0.499$ & $0.499\pi$ & $0.987\pi$ & $0.767\pi$ & $0.306\pi$ & 
$0.0177$ \\
      4 & $5$ & $0.499$ & $0$ & $0.499\pi$ & $0$ & $-0.785\pi$ & $0$ & 
$0.0158$
    \end{tabular}
  \end{ruledtabular}
  \caption{\label{tab:params} We list the binary parameters of the four precessing EOB waveforms that we consider in Sec.~\ref{sec:comparison}.  Case 3 corresponds to the NR simulation SXS:BBH:0052 of Ref.~\cite{Mroue:2013xna}, and case 4 corresponds to SXS:BBH:0058.} 
\end{table}
\begin{figure}
  \includegraphics[width=0.45\textwidth]{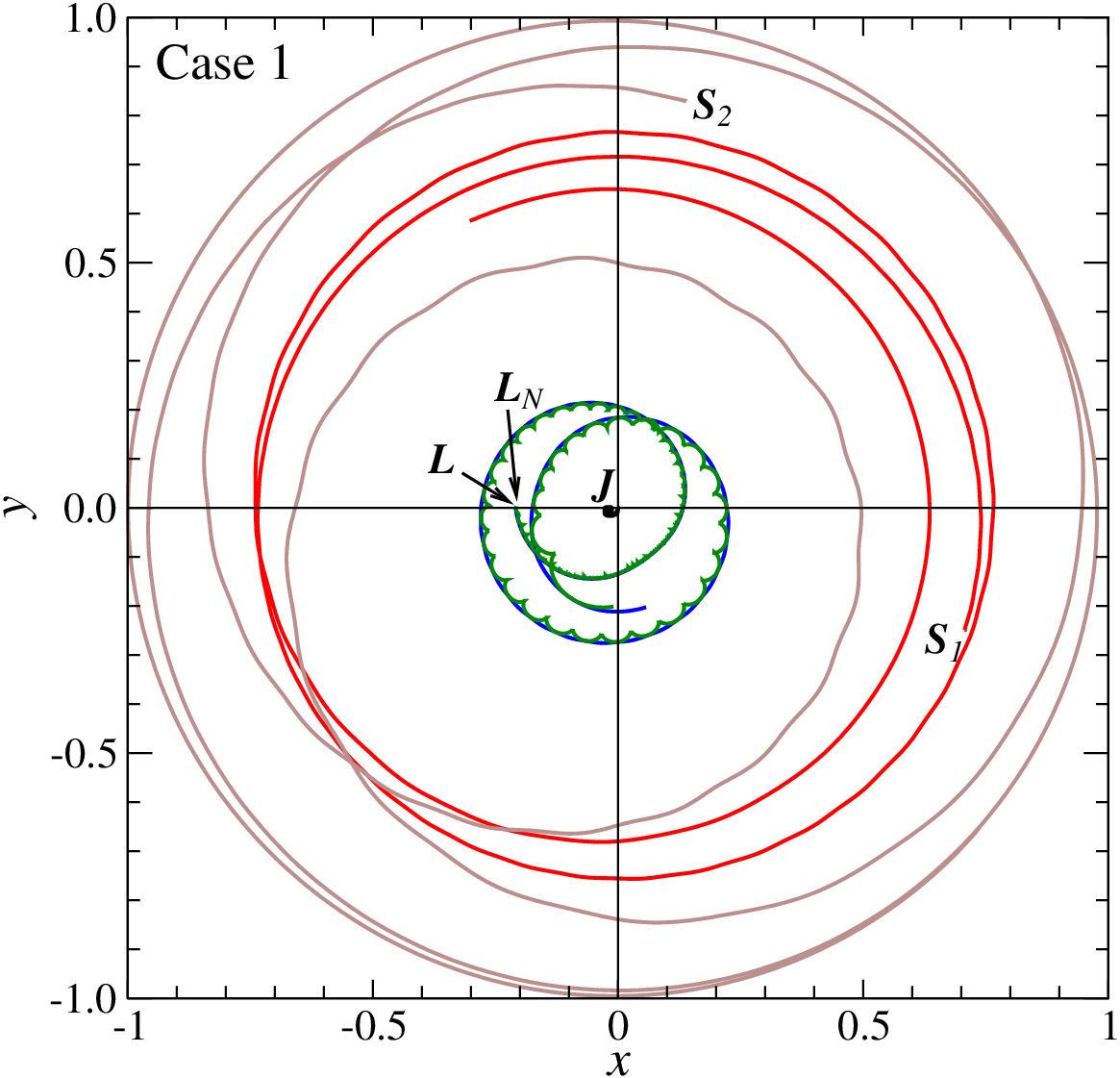}
  \vskip2ex
  \includegraphics[width=0.45\textwidth]{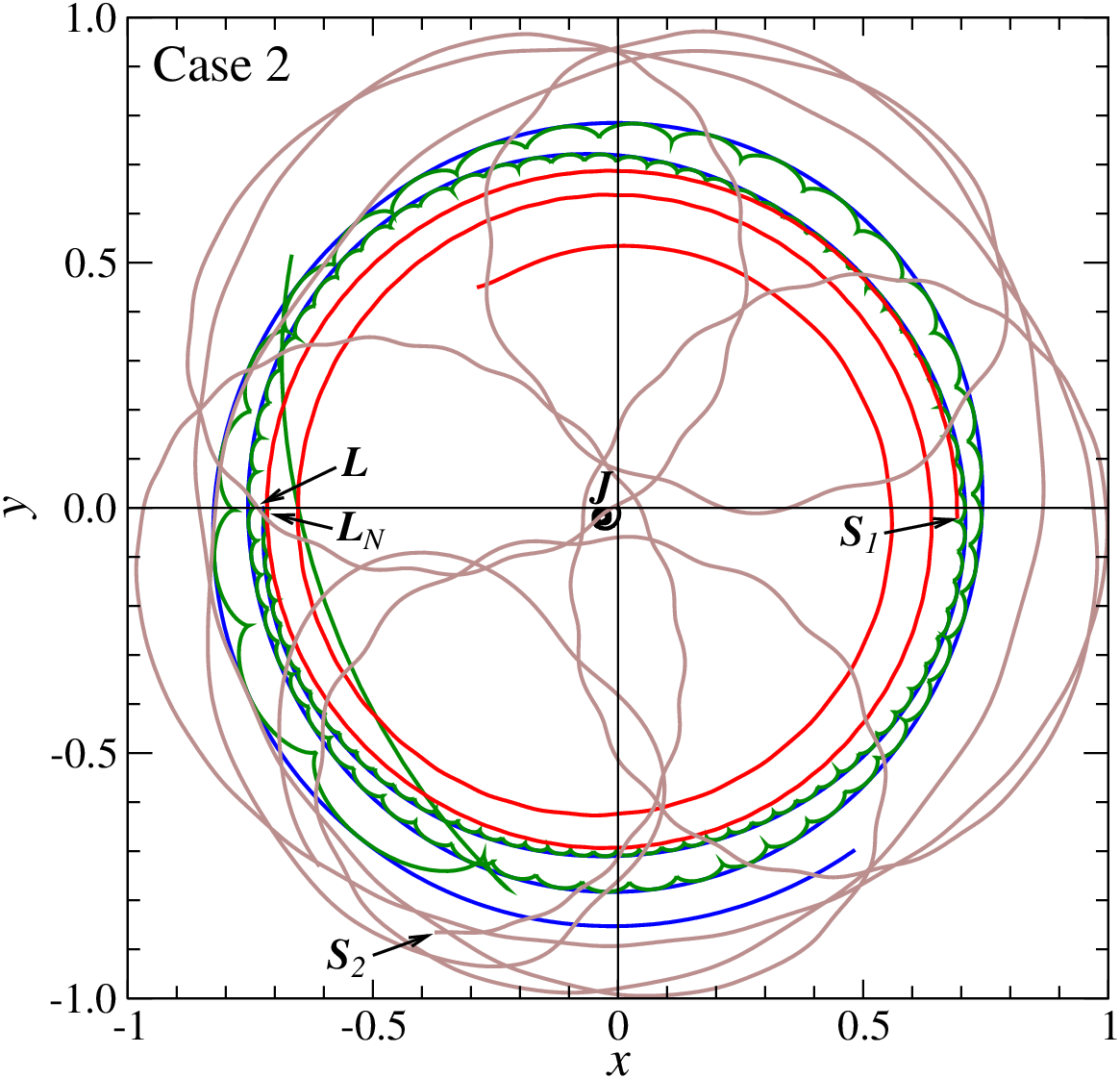}
  \caption{ \label{fig:traj} We show the projections of $\hat{\vJ}(t)$,
    $\hat{\vL}(t)$, $\hat{\vL}_N(t)$, $\hat{\vS}_1(t)$, and
    $\hat{\vS}_2(t)$ on the $x$-$y$ plane of the inertial frame whose
    $z$-axis is aligned with $\vJ(t^{\rm EOB}_{\rm \Omega peak})$. In
    the top and bottom panels we show trajectories of these unit
    vectors for cases 1 and 2 of Table~\ref{tab:params}, respectively. The initial point of each
    trajectory is marked by its name. The trajectory of $\hat{\vJ}(t)$
    ends at the origin, by definition. The trajectory of
    $\hat{\vL}_N(t)$ follows that of $\hat{\vL}(t)$ with oscillations
    due to nutation.}
\end{figure}

\section{Comparison between precessing waveforms}
\label{sec:comparison}

We generate four examples of EOB precessing waveforms using the model
defined in Sec.~\ref{sec:EOB}. The first two examples are a $q=2$ BH 
binary system exhibiting moderate precession-induced modulations and a $q=6$ 
binary system exhibiting strong modulations. The other two examples are chosen 
among the 171 NR simulations reported in Ref.~\cite{Mroue:2013xna} with the 
criterion of long and accurate waveforms exhibiting strong modulations. 
In these cases we compare NR, PN and EOB precessing waveforms. 
The physical parameters of the four binary configurations are listed in 
Table~\ref{tab:params}.

\subsection{Precessing and radiation-axis frames}
\label{sec:axis}

In Sec.~\ref{sec:p-frame}, we have proposed $\hat{\vL}_N(t)$ and
$\hat{\vL}(t)$ as possible basis vectors for the
precessing source frame. In this section, we compare their trajectories and
the corresponding precessing waveforms generated through their respective
precessing source frames. For convenience, we refer to waveforms generated in
these precessing source frames as the $\hat{\vL}_N(t)$-frame and $\hat{\vL}(t)$-frame 
waveforms, respectively. Furthermore, we extract the quadrupole-preferred radiation
axis~\cite{Schmidt2010} from the precessing waveforms and compare their trajectories 
with either $\hat{\vL}_N(t)$ or $\hat{\vL}(t)$.

\begin{figure*}
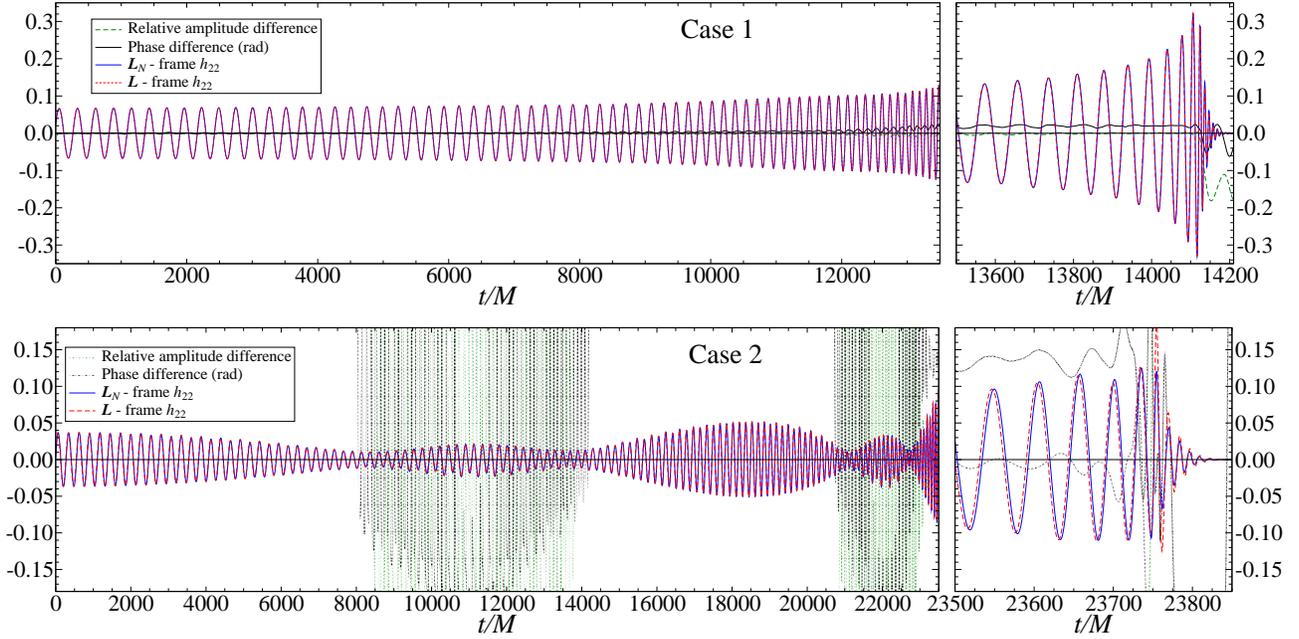

  \includegraphics[width=0.935\textwidth]{MildCaseWave} \\ \vspace*{0.2cm}
  \includegraphics[width=0.95\textwidth]{StrongCaseWave}
  \caption{ \label{fig:waveforms} We show the $\hat{\vL}_N(t)$-frame
    and the $\hat{\vL}(t)$-frame precessing waveforms, as well as
    their relative amplitude and phase differences. The top
    and bottom panels are waveforms for cases 1 and 2 of Table~\ref{tab:params}. 
    The left and right panels show
    the inspiral and the plunge-merger-ringdown stages of the
    waveforms, respectively.}
\end{figure*}

Figure~\ref{fig:traj} shows for cases 1 and 2 of Table~\ref{tab:params} 
the trajectories of the unit vectors $\hat{\vJ}$, 
$\hat{\vL}$, $\hat{\vL}_N$, $\hat{\vS}_1$, and $\hat{\vS}_2$ in the
plane perpendicular to $\vJ(t^{\rm EOB}_{\rm \Omega peak})$. In both
cases, the BHs complete more than two cycles of precession and the
directions of $\vJ(t)$ are well conserved during the entire inspiral
phase. All other vectors precess around $\vJ(t)$. These are expected
features of the well-known simple-precession picture of spinning
binaries in PN theory \cite{Apostolatos1994}. Another
common feature in both cases is the difference between the
trajectories of $\vL_N(t)$ and $\vL(t)$, which has been pointed out 
in Ref.~\cite{Ochsner:2012dj}. The trajectory of $\vL_{N}(t)$
shows nutation at twice the orbital frequency and its average follows
the smooth precession trajectory of $\vL(t)$. From PN theory~\cite{Kidder:1995zr}
\begin{equation}\label{LPN}
\vL=\vL_{N}+\vL_{\rm PN}+\vL_{\rm SO}+\mathcal{O}(c^{-4})\,,
\end{equation}
where 
\begin{eqnarray}
\vL_{\rm PN} &\equiv& \vL_{N}\left[\frac{1}{2}v^{2}(1-3\nu)+(3+\nu)\frac{M}{r}\right]\,,\\
\vL_{\rm SO} &\equiv& -\frac{2\mu }{r}\left[(\vS_\textrm{eff}\cdot \hat{\vL}_{N})\hat{\vL}_{N}+ (\vS_\textrm{eff}\cdot \hat{\vlambda})\hat{\vlambda}\right]\,,
\end{eqnarray}
with $v\equiv \hat{\Omega}^{1/3}$, $\hat{\vlambda} \equiv (\hat{\vL}_{N} \times \vvr) / r$ and
\begin{equation}
\vS_\textrm{eff}\equiv\left(1+\frac{3m_2}{4m_1} \right)\,\vS_1+\left(1+\frac{3m_1}{4m_2} \right)\,\vS_{2}\,.
\end{equation}
Note that the unit vector $\hat{\vlambda}$ instantaneously rotates about $\hat{\vL}_{N}$ at the orbital frequency $\Omega$. In addition, $\vL$ obeys a simple precession equation about $\vJ$, i.e. $\dot{\vL} \propto \vJ \times \vL$ (see Eq.~(2.13) of Ref.~\cite{Kidder:1995zr}). This, together with Eq.~(\ref{LPN}), implies that $\vL_{N}$ cannot simply precess about $\vJ$. When computing $\dot{\vL}_{N}$, the spin-orbit term $\vL_{\rm SO}$ generates contributions of the form 
\begin{equation}
(\vS_{\rm eff}\cdot \dot{\hat{\vlambda}})\hat{\vlambda}\quad\textrm{and}\quad(\vS_{\rm eff}\cdot\hat{\vlambda})\dot{\hat{\vlambda}}\,,
\end{equation}
which indeed oscillate at twice the orbital frequency, accounting for the nutations seen in Fig.~\ref{fig:traj}.

The main difference between the two cases is the size of
the opening angle between $\vJ(t)$ and $\vL(t)$ and correspondingly
the strength of the orbital precession. In the comparable-mass $q=2$ case,
$\vL(t)$ always dominates over the BH spins during inspiral and the
angle between $\vJ(t)$ and $\vL(t)$ remains small. The orbital
precession is therefore mild. In the $q=6$ case, on the contrary, the
contribution of $\vS_1(t)$ to $\vJ(t)$ is comparable to that of
$\vL(t)$ initially and becomes more and more dominant. Because of the
large opening angle between $\vJ(t)$ and $\vL(t)$, the direction of
$\vL(t)$ changes more than $\pi/2$ during precession and an 
initially face-on binary becomes edge-on a few times during the inspiral.

\begin{figure}
\includegraphics[width=0.45\textwidth]{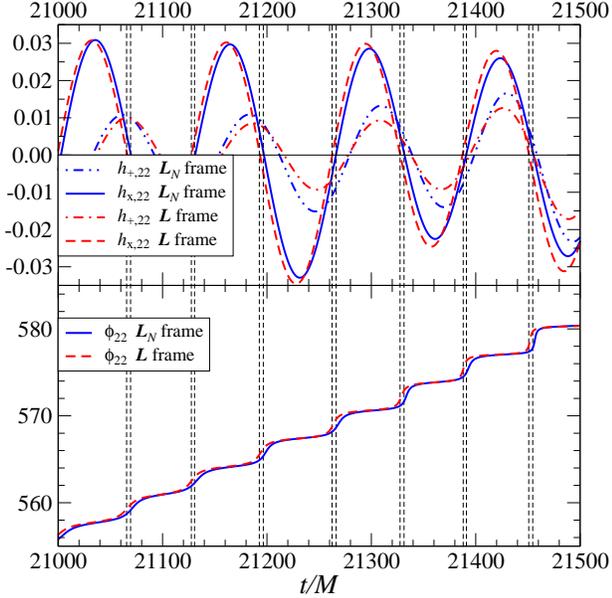}
\caption{\label{fig:phaseburst} For case 2 of Table~\ref{tab:params},
  we show the $\hat{\vL}_N(t)$-frame and $\hat{\vL}(t)$-frame
  waveforms in the top panel and their phase evolutions in the bottom
  panel, over a short time period from $t=21\,000M$ to $21\,500M$. The
  vertical lines mark the time when the dominant quadrature (the
  imaginary part for this specific instance) of any waveform becomes
zero. It coincides with the time when the corresponding phase
evolution in the bottom panel experiences a rapid growth. The absolute
phase values are not relevant.}
\end{figure}

In Fig.~\ref{fig:waveforms}, we compare precessing waveforms generated
in the $\hat{\vL}_N(t)$ and $\hat{\vL}(t)$ precessing source 
frames. Considering the oscillatory difference between the trajectories of
$\hat{\vL}_N(t)$ and $\hat{\vL}(t)$ shown in Fig.~\ref{fig:traj}, it
is somewhat unexpected that the precessing waveforms agree quite
well. In case 1, the waveforms are visually indistinguishable
during inspiral --- with relative amplitude difference below $1\%$ and
phase difference below 0.02 radians. Even in the $q=6$ case 2, where precession is strong, the waveforms agree reasonably well. Although the
relative amplitude and phase differences oscillate strongly when
the amplitudes of the waveforms are small, their averages differ only by $< 5\%$ and
$< 0.15$ radians over the $\sim 24\,000 M$ long inspiral. The
oscillations are due to the precession-induced modulation and are
expected to be strong when the orbital plane goes through a nearly
edge-on phase, corresponding to small waveform amplitudes.  

In Fig.~\ref{fig:phaseburst}, we examine closely the waveforms as well as
their phase evolutions over a time period of $500M$. The real and
imaginary parts of $h_{22}$, i.e. its $+$ and $\times$ polarizations
in the radiation frame, show substantial amplitude difference, implying a
deviation from circular polarization due to the orbital plane
inclination. The phase evolves most rapidly when the dominant 
quadrature (the imaginary part in this example) goes through zero. Even a small difference in the times
when this happens for the two waveforms leads to a burst of phase
difference. Such phase differences can be partly removed by 
time-shifting the two waveforms, but not through a phase shift.
In spite of these bursts of amplitude and phase difference, the overall
agreement of the waveforms is good. The overlaps between the 
$\hat{\vL}_N(t)$-frame and $\hat{\vL}(t)$-frame waveforms, optimized
over time and phase of coalescence, are above 0.999 in case 1 and above 0.985 in 
case 2~\footnote{The overlaps are calculated using the zero-detuned high-power Advanced LIGO noise
  curve \cite{Shoemaker2009} for the range of binary total masses from
  $20$ to $200M_\odot$.}. The lower overlaps in case 2 are due to the larger 
difference between $\hat{\vL}_N(t)$-frame and $\hat{\vL}(t)$-frame waveforms during merger and 
ringdown.

Finally, we examine the preferred radiation axis determined by the waveforms 
extracted at infinity. Since we developed only the $\ell=2$ modes in the 
current EOB model, we calculate the
quadrupole-preferred radiation axis~\cite{Schmidt2010} with a small
modification. In Ref.~\cite{Schmidt2010}, the quadrupole-preferred
axis is determined by maximizing the power in the $(2,2)$ and $(2,-2)$
modes of the Weyl scalar $\Psi_4(t)$. We determine the
quadrupole-preferred axis by maximizing the power in the strain modes
$h_{22}(t)$ and $h_{2,-2}(t)$. Specifically, given the $h^{(o)}_{2m}(t)$ modes in an
arbitrary original frame, the $h^{(n)}_{2m}(t)$ modes in a new frame are given
by Eq.~\eqref{rotation}; so we compute the Euler angles
$\alpha(t)$, $\beta(t)$ and $\gamma(t)$ (defining the rotation from the 
original to the new frame) that maximize the quantity 
$|h^{(n)}_{22}(t)|^2+|h^{(n)}_{2,-2}(t)|^2$. The quadrupole-preferred axis is then
given by the $z$-axis of the new frame defined by these Euler
angles. We find that the quadrupole-preferred axis computed from $\hat{\vL}_N(t)$-frame or
$\hat{\vL}(t)$-frame waveforms agrees with $\hat{\vL}_N(t)$ or $\hat{\vL}(t)$ to within $0.3^\circ$ 
during inspiral. That is to say, the preferred radiation axis determined by EOB
waveforms coincides with the reference axis ($\hat{\vL}_N(t)$ or $\hat{\vL}(t)$) 
of the precessing frame determined by the EOB dynamics. Therefore, comparisons 
of preferred radiation axes determined by NR and EOB precessing waveforms will 
provide direct information for calibrating the precession dynamics, in particular the dynamics 
of $\hat{\vL}_N(t)$ and $\hat{\vL}(t)$, of the EOB model.  
\begin{figure}
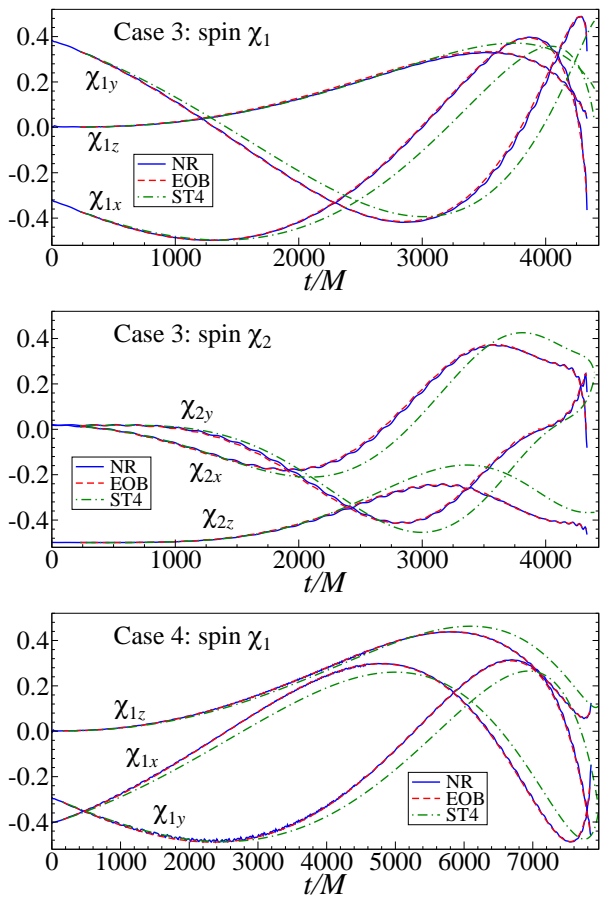

  \includegraphics[width=0.44\textwidth]{SpECase52_spin1}   \\ \vspace{0.2cm}
  \includegraphics[width=0.44\textwidth]{SpECase52_spin2}   \\ \vspace{0.2cm}
  \includegraphics[width=0.44\textwidth]{SpECase59_spin1}
  \caption{ \label{fig:NRspins} We show evolutions of the dimensionless spin vectors $\vchi_1=\vS_1(t)/m_1^2$ and $\vchi_2=\vS_2(t)/m_2^2$ of the NR simulation and the EOB and ST4 models. Specifically, we show the projections of $\vchi_1$ and $\vchi_2$ on the basis vectors of the inertial source frame $\{\ve_1^S,\ve_2^S,\ve_3^S\}$ that is aligned with the initial orbital orientation $[\hat{\vL}_N]_0$ (see Fig.~\ref{fig:frames}). The top two panels show $\vchi_1$ and $\vchi_2$ for case 3 of Table~\ref{tab:params}. The bottom panel shows $\vchi_1$ ($\vchi_{2}=0$) for case 4 of Table~\ref{tab:params}. The EOB and ST4 data start at the after-junk-radiation time in the NR simulations, which are $t=230M$ and $t=160M$ for cases 3 and 4, respectively.} 
\end{figure}
\begin{figure*}
  \includegraphics[width=0.95\textwidth]{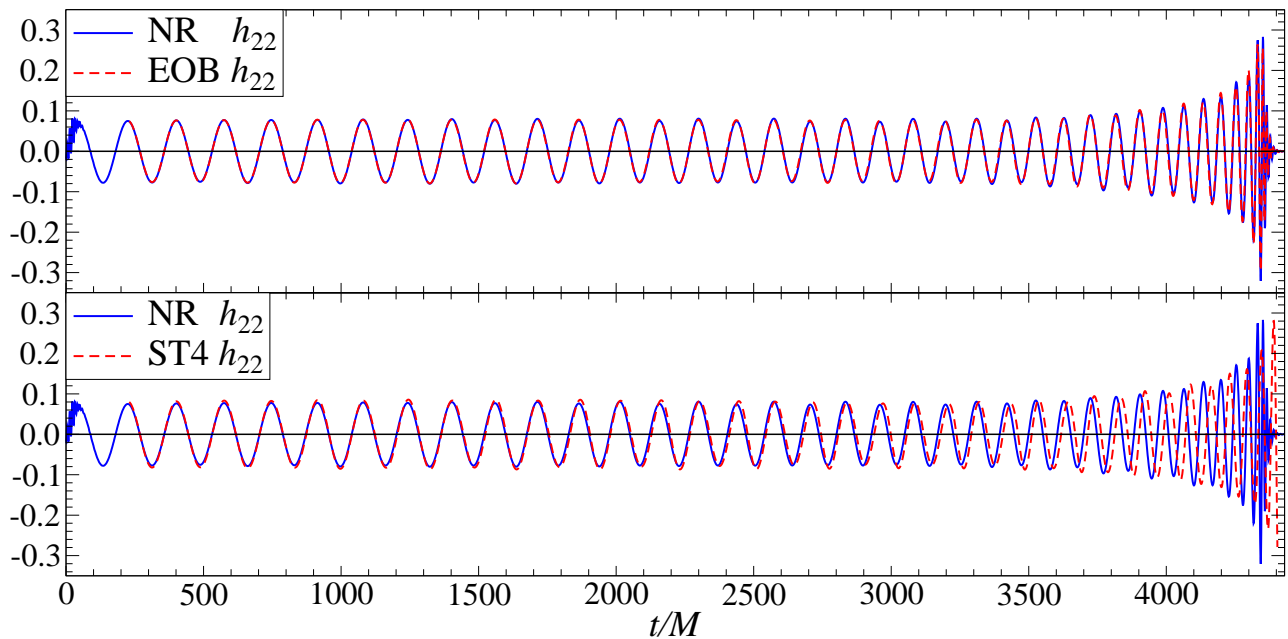}
  \caption{ \label{fig:NRwaves1} We show for case 3 of Table~\ref{tab:params} the $h_{22}$ mode decomposed in the inertial source frame $\{\ve_1^S,\ve_2^S,\ve_3^S\}$ that is aligned with the initial orbital orientation $[\hat{\vL}_N]_0$ (see Fig.~\ref{fig:frames}). For clarity, we show the NR and EOB $h_{22}$ in the top panel and the NR and ST4 $h_{22}$ in the bottom panel. The EOB and ST4 data start at the after-junk-radiation time of $t=230M$.} 
\end{figure*}
\begin{figure*}
  \includegraphics[width=0.95\textwidth]{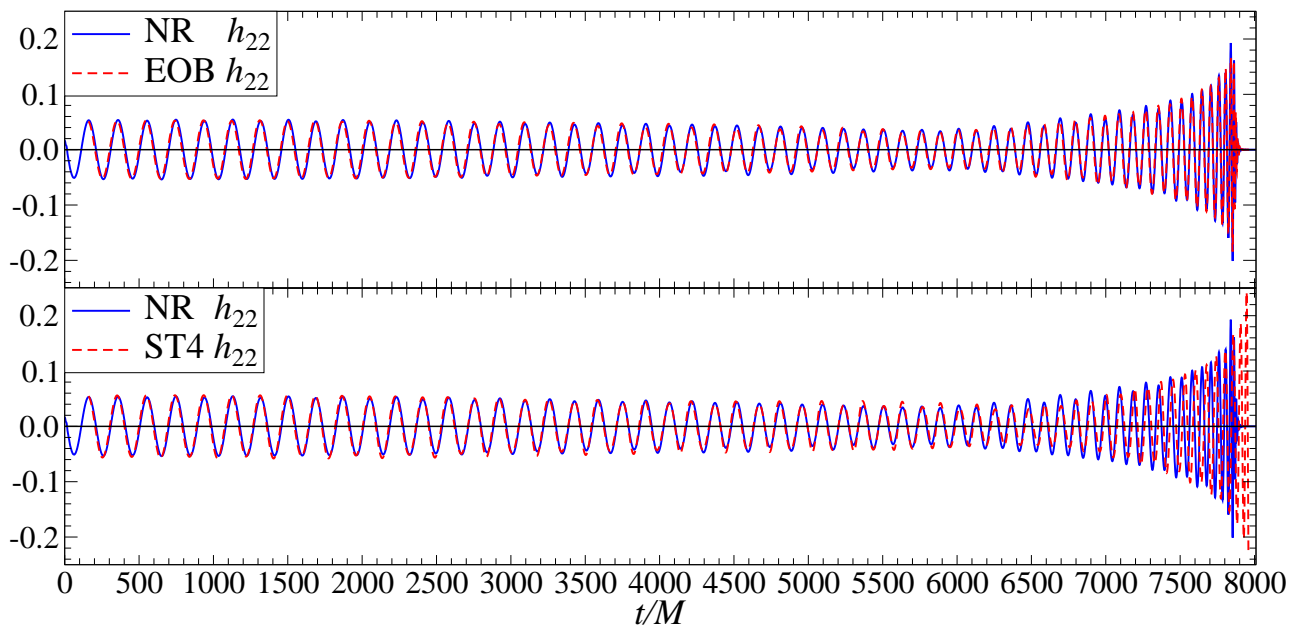}
  \caption{ \label{fig:NRwaves2} We show for case 4 of Table~\ref{tab:params} the $h_{22}$ mode decomposed in the inertial source frame $\{\ve_1^S,\ve_2^S,\ve_3^S\}$ that is aligned with the initial orbital orientation $[\hat{\vL}_N]_0$ (see Fig.~\ref{fig:frames}). For clarity, we show the NR and EOB $h_{22}$ in the top panel and the NR and ST4 $h_{22}$ in the bottom panel. The EOB and ST4 data start at the after-junk-radiation time of $t=160M$.} 
\end{figure*}

\subsection{Comparison with numerical-relativity waveforms}
\label{sec:EOBvsNR}

The precessing EOB model defined in Sec.~\ref{sec:EOB} is not calibrated to any 
precessing numerical simulations. The only nonperturbative information extracted from 
NR simulations and employed in this precessing EOB model is contained in the spinning, 
nonprecessing sector, which was calibrated to {\it only} two equal-mass, spinning, 
nonprecessing numerical simulations~\cite{Chu2009} and five nonspinning ones\cite{Buchman:2012,Scheel2009} in Ref.~\cite{Taracchini2012}. 
It is therefore highly interesting to compare the EOB precessing waveforms to NR waveforms. 

The Caltech-Cornell-CITA collaboration has recently produced a large number of long and accurate 
waveforms~\cite{Mroue:2013xna}. We choose among them two precessing waveforms that are sufficiently long 
($\sim 35$ and $\sim 65$ GW cycles) and display strong precessional modulations. The physical parameters of 
these two waveforms are given in the last two rows of Table~\ref{tab:params}. We compare those numerical 
waveforms also with the PN SpinTaylorT4 (ST4) inspiraling waveforms~\cite{Buonanno:2002fy}, which are commonly 
used in the literature and in LIGO and Virgo software. We generate the ST4 waveforms at the highest PN 
order available today, namely spin-amplitude corrections through 1.5PN 
order~\cite{Arun:2008kb}~\footnote{The 2PN spin-amplitude corrections have been derived in 
Ref.~\cite{Buonanno:2012rv}. Since they are not yet implemented in any ready-to-use software package 
and are not crucial for the purpose of our comparisons, we do not include them here.}
and phase corrections through 3.5PN order~\cite{Bohe:2013cla} using the LIGO Algorithm Library~\cite{LAL}.

We extract the initial values of $\vS_1$, $\vS_2$ and GW frequency from the NR data soon after the 
junk radiation, which typically carries away unphysical radiation present in the initial data. We then 
set EOB and ST4 initial conditions using these values and start their evolutions after the 
junk-radiation time, which is $t=230M$ for case 3 of Table~\ref{tab:params} and 
$t=160M$ for case 4 of Table~\ref{tab:params}. We align the orbital orientation $\hat{\vL}_N$ 
at these after-junk-radiation times with the inertial 
source frame $\{\ve_1^S,\ve_2^S,\ve_3^S\}$ (see Fig.~\ref{fig:frames}) and use it as the default 
frame for our comparisons. Unlike the case of nonprecessing dynamics and waveforms, we must impose specific 
$\vS_1$ and $\vS_2$ directions relative to the initial binary separation vector $\vvr_0$ at a 
specific after-junk-radiation time. Thus, we {\it do not} apply any time or phase shifts when comparing 
numerical and analytical waveforms. 

In Fig.~\ref{fig:NRspins} we compare the evolutions of the dimensionless
spin vectors $\vchi_1(t)=\vS_1(t)/m_1^2$ and $\vchi_2(t)=\vS_2(t)/m_2^2$ ($\vchi_2=0$ for case 4) 
for the NR, EOB and ST4 dynamics. Quite remarkably, the EOB spins follow the NR ones rather accurately
all the way through the inspiral-plunge stage, while the ST4
spins, although capturing the qualitative precessional behavior of the
NR ones, show quantitative differences in both the inspiral and
precessional time scales. 

\begin{figure*}
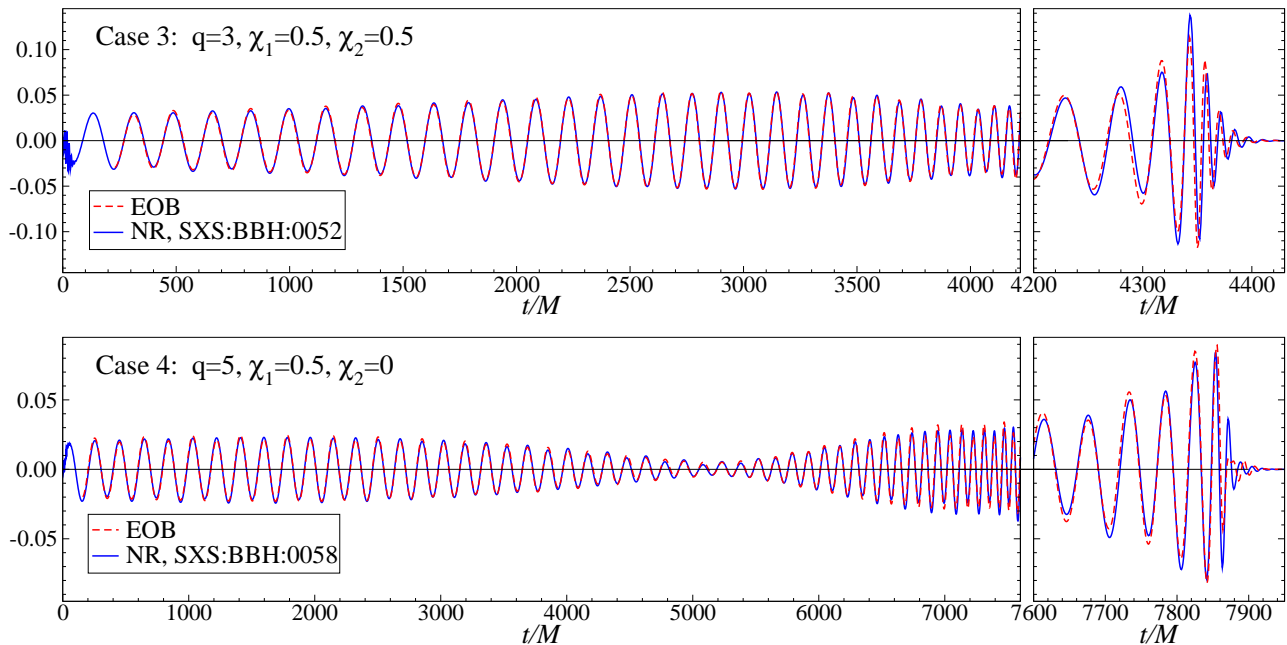

  \includegraphics[width=0.95\textwidth]{SpECase52_hp}   \\ \vspace{0.2cm}
  \includegraphics[width=0.95\textwidth]{SpECase59_hp}
  \caption{ \label{fig:NRwaves_hp} We show for cases 3 and 4 of
    Table~\ref{tab:params} the GW polarization $h_+$, containing 
    contributions from $\ell=2$ modes, that propagates
    along a direction $\hat{\vN}$ specified by spherical coordinates
    $\theta=\pi/3$ and $\phi=\pi/2$ associated with the inertial
    source frame $\{\ve_1^S,\ve_2^S,\ve_3^S\}$. The EOB waveforms
    start at the after-junk-radiation times of $t=230M$ and $t=160M$,
    respectively.}
\end{figure*}

In Figs.~\ref{fig:NRwaves1} and \ref{fig:NRwaves2}, we compare NR, EOB
and ST4 $h_{22}$ modes decomposed in the inertial source frame
$\{\ve_1^S,\ve_2^S,\ve_3^S\}$. Since the source frame is aligned with
the initial orbital orientation $[\hat{\vL}_N]_0$ and the binary orbit
precesses only moderately in case 3, there are only moderate
modulations on $h_{22}$ in this case. The modulations in case 4 are
strong, though. In both cases, the agreement between NR and EOB
$h_{22}$ modes is remarkable. Their amplitudes agree quite well and
their phases, aligned at the initial time, differ by only $\sim 0.2$
rad at merger, i.e., at the peak of the NR (2,2) mode.  The agreement
between NR and ST4 $h_{22}$ modes, although not comparable with the
agreement between NR and EOB, is also very good. Even though the
amplitudes differ by $\sim 10\%$ during the inspiral, because
amplitude corrections are known only through 1.5PN order in the
spinning case~\cite{Arun:2008kb}, their phases agree quite well for
tens of cycles but start departing from each other $10$ GW cycles
before merger. Quite interestingly, we have found that using the newly
available 3.5PN spin-orbit effects~\cite{Bohe:2013cla} in the phasing
of ST4 improves the agreement with the NR waveforms. If we were using
the 2.5PN phasing, the end of the inspiral would occur $\sim 460 M$
$(960 M)$ instead of $\sim 60 M$ $(140 M)$ after the merger of the NR
waveform, for case 3 (4). Moreover, we find for cases 3 and 4 that when we align 
the 3.5PN and NR phasing at the after--junk-
radiation time, they accumulate a difference of $1$ GW cycle only $1$ 
cycle before merger. By contrast the 2.5PN phasing differs from the NR phasing 
by $1$ GW after $28$ ($37$) GW cycles [or  $6$ ($16$) GW cycles before merger] for
  case 3 (4).

The agreement between NR and EOB modes $(2,1)$ and $(2,0)$ modes are also very good.
Rather than the modes, we show in Fig.~\ref{fig:NRwaves_hp} the NR and EOB polarizations $h_+(t)$ given by
Eq.~\eqref{hphc}. Since only the $\ell=2$ modes are available in the current precessing EOB model,
we limit the summation over $\ell$ to only $\ell=2$. To include
substantial contributions from all $\ell=2$ modes, we choose
$\theta=\pi/3$ and $\phi=\pi/2$ for the direction of GW propagation
$\vN$ (see Fig.~\ref{fig:frames}). As expected from the very good
agreement of the individual modes, the NR and EOB polarizations also
agree remarkably. 

\begin{figure}
  \includegraphics[width=0.39\textwidth]{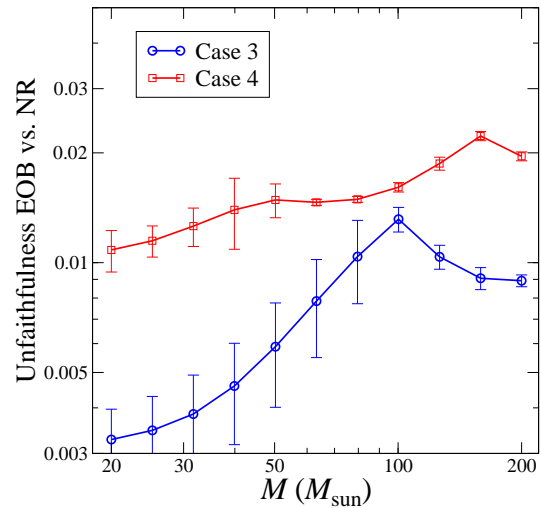}
  \caption{ \label{fig:unfaithfulness} Unfaithfulness of the EOB waveforms when compared to NR waveforms as a function of the total binary mass.   Shown are cases 3 and 4 of Table~\ref{tab:params}. 
The error bars are estimates of numerical errors. The direction of GW propagation $\hat{\vN}$ is specified 
by the spherical coordinates $\theta=\pi/3$ and $\phi=\pi/2$.}
\end{figure}

Finally, we measure the difference between EOB and NR polarizations
with the {\it unfaithfulness}~\cite{Damour98}, defined as 
\begin{equation}\label{unF}
\bar{\mathcal{F}}=1-\max_{t_c,\phi_c,\psi}\frac{\langle h_{\rm NR}|h_{\rm EOB}\rangle}{\sqrt{\langle h_{\rm NR}|h_{\rm NR}\rangle \langle h_{\rm EOB}|h_{\rm EOB}\rangle}} \,,
\end{equation}
where the EOB waveform of the detector response is
\begin{eqnarray}
h_{\rm EOB}(t;t_c,\phi_c,\psi,\vlambda)\propto &\cos\psi& h_{{\rm EOB,}+}(t;t_c,\phi_c,\vlambda) \nonumber \\
+&\sin\psi& h_{{\rm EOB,}\times}(t;t_c,\phi_c,\vlambda)\,, \nonumber \\
\end{eqnarray}
and the maximization is over the time and phase of coalescence $t_c$
and $\phi_c$, as well as the polarization angle $\psi$ that combines
the $+$ and $\times$ polarizations in the radiation frame. We do not 
optimize over the physical binary parameters $\vlambda$, i.e.,we use 
the same $\vlambda$ in $h_{\rm NR}$ and $h_{\rm EOB}$. Note that since
we include modes with different $m$, $\phi_c$ and $\psi$ are no
longer degenerate and both of them have to be maximized over.  We
define the inner product between two waveforms through the following
integral in the frequency domain
\begin{equation}
\langle h_1,h_2\rangle\equiv 4{\rm Re}\int_0^\infty\frac{\tilde{h}_1(f)\tilde{h}_2^*(f)}{S_h(f)}df \,,
\end{equation}
where $\tilde{h}_1(f)$ and $\tilde{h}_2(f)$ are frequency domain
waveforms and $S_h(f)$ is the noise power spectral density of the
detector. We employ the zero-detuned high-power advanced LIGO noise
curve \texttt{ZERO\_DET\_HIGH\_P} given in~\cite{Shoemaker2009}.  The
NR waveforms, although very long, cover the entire advanced LIGO frequency
band only for $M\ge 100M_\odot$.  Thus, to reduce artifacts when
considering binaries with $M <100M_\odot$, we taper both ends of the
NR and EOB waveforms using the Planck-taper window
function~\cite{McKechan:2010kp} (see Ref.~\cite{PanEtAl:2011} for
details).  In Fig.~\ref{fig:unfaithfulness}, we show the EOB
unfaithfulness when the total mass $M$ varies between $20 M_\odot$ and
$200M_\odot$. We choose the same direction of GW propagation
  $\vN$ as is considered in Fig.~\ref{fig:NRwaves_hp}, namely
  $\theta=\pi/3$ and $\phi=\pi/2$.

For each waveform we estimate the numerical error in the unfaithfulness results 
of Fig.~\ref{fig:unfaithfulness} by calculating the unfaithfulness 
of the EOB waveform with two numerical waveforms: the extrapolated high-resolution waveform shown 
in Fig.~\ref{fig:NRwaves_hp} and the outermost finite-radius high-resolution waveform. 
We use the difference between these unfaithfulness results to estimate the extrapolation error.
We might estimate the finite resolution errors in the same way 
by calculating the unfaithfulness of the EOB waveform with the extrapolated 
high- and medium-resolution numerical waveforms. However, medium resolution 
simulations for these two cases are not available, but we expect from 
previous studies that errors due to resolution are smaller than 
errors due to extrapolation~\cite{PanEtAl:2011}.

Since the unfaithfulness of EOB waveforms is below $\sim2\%$, we expect that 
the ineffectualness, which measures the difference between EOB and NR waveforms when 
minimizing also over the binary parameters $\vlambda$, will be below $1\%$ (with a 
loss of event rates less than $3\%$). Thus, for those two precessing binary 
configurations, the EOB waveforms are sufficiently accurate for detection with 
advanced LIGO detectors. 

Although these very encouraging results refer only to two precessing 
binary configurations, they strongly suggest that the approach we have proposed 
for modeling precessing compact binaries within the EOB model is bound to succeed. 
A more comprehensive and careful comparison of the EOB model with a larger number 
of accurate NR simulations will be carried out in the near future using the entire catalog 
of simulations in Ref.~\cite{Mroue:2013xna}.

\section{Conclusions}
\label{sec:conclusions}

So far, the EOB modeling of GWs emitted from compact binaries has
focused primarily on nonprecessing binary
configurations~\cite{Buonanno2007, DN2007b,
  DN2008,Buonanno:2009qa,Damour2009a, Pan:2009wj, PanEtAl:2011,
  Taracchini2012,Damour:2012ky}. Nonspinning EOB waveforms have been
employed in the first searches of GWs from high-mass binary BHs with
LIGO and Virgo
detectors~\cite{Abadie:2011,Aasi:2012rja,Abadie:2012aa}.  Recently,
studies carried out within the NRAR collaboration~\cite{Hinder:2013oqa}
have shown that nonprecessing EOB waveforms originally calibrated to 
seven NR waveforms~\cite{Chu2009,Scheel2009,Buchman:2012} in
Ref.~\cite{Taracchini2012} match very well also tens of new NR
waveforms produced within the NRAR collaboration.  The next,
challenging task is to achieve a similar success also for generic,
spinning binary configurations. In this paper we have started
addressing this important problem.

Building on previous work~\cite{Buonanno:2002fy,Buonanno:2005xu,Barausse:2009aa,Barausse:2009xi,Barausse:2011ys,
Pan:2009wj,Taracchini2012}, we have proposed a strategy to generate EOB precessing waveforms. The procedure 
employs the precessing convention of Ref.~\cite{Buonanno:2002fy} that minimizes the precession-induced 
modulations in the waveform's phase and amplitude, and an inertial frame aligned with the spin of 
the final BH where the matching between the inspiral-plunge and  merger-ringdown EOB waveforms is carried out. 

When spins are aligned or antialigned with the orbital angular momentum, the EOB precessing waveforms 
that we have built reduce to the nonprecessing EOB waveforms calibrated to seven 
nonprecessing NR waveforms 
in Ref.~\cite{Taracchini2012}. Since the factorized energy flux is not yet available for 
precessing spins, we have included in the radiation-reaction force of the EOB 
dynamics only spin couplings whose projection along the orbital angular momentum is different 
from zero. This limitation will be relaxed in the future as soon as the radiation-reaction sector of the 
EOB model is improved. Furthermore, we have limited this first study to the EOB $\ell =2$ modes.

Without recalibrating the EOB precessing waveforms, we have then compared them to two, long, strongly 
precessing NR waveforms that were recently produced in Ref.~\cite{Mroue:2013xna}. We have found a remarkable 
agreement both for the dynamics, that is the spins' components, and the gravitational polarizations. 
In particular, when using the advanced-LIGO noise spectral density, the mismatches between 
the EOB and NR waveforms for binary masses $20\mbox{--}200 M_\odot$ are below $2\%$ when 
maximizing only on the time and phase at coalescence and on the polarization angle. Although 
those results only refer to two binary configurations, they are very encouraging and suggest 
that the EOB precessing model developed here is an excellent starting point for building a generic, 
spinning EOB model for advanced LIGO and Virgo searches. We have
 also compared the 
two NR waveforms to PN ST4 waveforms that are largely used in the literature and 
in LIGO and Virgo software. We have found that the PN waveforms at 3.5PN order agree very well 
with NR waveforms for several GW cycles, and accumulate a phase difference of $\sim 6$ rad, 
starting about 10 GW cycles before merger. 

Finally, several analyses were left out in this first study of precessing waveforms. They include 
(i) a more detailed comparison between spin variables in the numerical simulations and analytical 
models, (ii) the extension of precessing waveforms to modes higher than $\ell=2$, (iii) a more 
systematic way of identifying the initial conditions in the numerical and analytical waveforms, 
and (iv) the inclusion of resolution errors when estimating numerical errors. 
We defer those important extensions to a future publication where many more NR waveforms will be also 
analysed.

\begin{acknowledgments} 
We thank An\i l Zengino\u{g}lu, Geoffrey Lovelace and Mike Boyle for their contributions
to the productions of the NR waveforms used in this paper.
A.B., Y.P. and A.T. acknowledge partial support from NSF Grants
No. PHY-0903631 and No. PHY-1208881.  A.B. also acknowledges partial
support from the NASA Grant NNX12AN10G. 
A.M. and H.P. gratefully acknowledge support from NSERC of Canada, the Canada Chairs Program, 
and the Canadian Institute for Advanced Research.
L.K. gratefully acknowledges support from the
Sherman Fairchild Foundation, and from NSF grants PHY-0969111 and
PHY-1005426.  
Simulations used in this work
were computed with the \texttt{SpEC} code~\cite{SpECwebsite}.  Computations
were performed on the Zwicky cluster at Caltech, which is supported by
the Sherman Fairchild Foundation and by NSF award PHY-0960291; on the
NSF XSEDE network under grant TG-PHY990007N; and on the GPC
supercomputer at the SciNet HPC Consortium~\cite{scinet}. SciNet is
funded by: the Canada Foundation for Innovation under the auspices of
Compute Canada; the Government of Ontario; Ontario Research
Fund--Research Excellence; and the University of Toronto.
\end{acknowledgments}


\end{document}